\UseRawInputEncoding

\documentclass[aps,pra,twocolumn,showpacs,superscriptaddress]{revtex4-1}
\usepackage{graphicx}
\usepackage{dcolumn}
\usepackage{bm}
\usepackage{color}
\usepackage{ulem}
\usepackage{times}
\usepackage{amssymb}
\usepackage{amsmath}
\usepackage{epsfig}
\usepackage{epstopdf}
\usepackage{dsfont}
\usepackage{subfigure}
\usepackage[colorlinks=true,citecolor=blue, linkcolor=blue,urlcolor=blue]{hyperref}
\usepackage{setspace}
\usepackage{bookmark}
\usepackage{color}
\usepackage{subfigure}

\newenvironment{proof}{{\indent  \it Proof:\,}}{\hfill $\blacksquare$\par}
\definecolor{Green2}{rgb}{0.78,0.93,0.8}
\definecolor{Green3}{rgb}{0.80,0.87,0.76}
\definecolor{Green1}{rgb}{0.77,0.88,0.70}
\begin{document}


\title{Complementary relations of entanglement, coherence, steering
and Bell nonlocality inequality violation in \textcolor[rgb]{0.00,0.00,0.00}{three-qubit states}}


\author{Dong-Dong Dong}
\affiliation{School of Physics and Optoelectronics Engineering, Anhui University, Hefei
  230601,  People's Republic of China}
\author{Xue-Ke Song}%
\email{songxk@ahu.edu.cn}
\affiliation{School of Physics and Optoelectronics Engineering, Anhui University, Hefei
  230601,  People's Republic of China}
\author{Xiao-Gang Fan}
\affiliation{School of Physics and Optoelectronics Engineering, Anhui University, Hefei
  230601,  People's Republic of China}

\author{Liu Ye}
\affiliation{School of Physics and Optoelectronics Engineering, Anhui University, Hefei
  230601,  People's Republic of China}
\author{Dong Wang}
\email{dwang@ahu.edu.cn}
\affiliation{School of Physics and Optoelectronics Engineering, Anhui University, Hefei
  230601,  People's Republic of China}


\date{\today}

\begin{abstract}
  We put forward complementary relations of entanglement, coherence, steering inequality violation, \textcolor[rgb]{0.00,0.00,0.00}{and Bell nonlocality}
  for arbitrary three-qubit states.
  We show that two families of genuinely entangled three-qubit pure states with single parameter exist,
  and they exhibit maximum coherence and steering inequality violation for a fixed amount of negativity,
  respectively. It is found that the negativity is exactly equal to the geometric mean of bipartite
  concurrences for the three-qubit pure states, although the negativity is always less than or equal
  to the latter for three-qubit mixed states. Moreover, the complementary relation between negativity
  and first-order coherence for tripartite entanglement states are established. Furthermore, we
  investigate the close relation between the negativity and the maximum steering inequality violation.
  \textcolor[rgb]{0.00,0.00,0.00}{
    In addition, the complementary relation between negativity and the maximum Bell-inequality violation for arbitrary three-qubit states is obtained.
    The results provide reliable evidence of fundamental connections among entanglement, coherence,
    steering inequality violation, and Bell nonlocality.}
\end{abstract}


\maketitle

\section{introduction}

Entanglement is considered to
be the most fundamental features of quantum mechanics. It has many potential applications in quantum information processing, including canonical ones: quantum cryptography \cite{PhysRevLett.67.661}, quantum teleportation \cite{PhysRevLett.70.1895}, and
dense coding \cite{PhysRevLett.69.2881}.  
It occurs when state of a composite system cannot be written as a product of states of each subsystem. The information-theoretic quantification of entanglement is connected with its usefulness in terms of quantum computing and communication. There are a number of measures for quantifying the bipartite or multipartite entanglement, such as the concurrence \cite{PhysRevLett.80.2245}, entanglement of formation \cite{doi:10.1080/09500349908231260}, negativity \cite{PhysRevA.65.032314,sabin2008classification}, the geometric mean of bipartite concurrences (GBC) \cite{PhysRevResearch.4.023059,PhysRevA.71.012318}, and so on. Although these measures are regarded differently, there exists many evidences to show that they are in fact strongly related \cite{PhysRevA.70.032326,miranowicz2004comparative,PhysRevA.91.032327}, and even potentially equivalent \cite{PhysRevA.106.042415}. For example, Wootters found a functional relation between the entanglement of formation and concurrence \cite{PhysRevLett.80.2245}.

Coherence, directly related to interference phenomena, describes the coherent superposition of states of interaction fields \cite{nielsen_chuang_2010,RevModPhys.89.041003}. It has been regarded as a useful physical resource possessing different computable measures, such as $l_1$ norm \cite{PhysRevLett.113.140401}, relative entropy \cite{PhysRevLett.113.140401,PhysRevLett.116.120404}, and skew information \cite{PhysRevLett.113.170401}. It plays an important role in \textcolor[rgb]{0.00,0.00,0.00}{quantum thermodynamics \cite{PhysRevLett.111.250404} and witnessing quantum correlations} \cite{hu2016extracting}.
In addition, quantum steering can be captured as another effective quantum resource with local operations assisted by one-way classical communication as the free operations \cite{PhysRevX.5.041008}.
It shows a special phenomenon of quantum information that the correlation of a two-particle state allows one to steer the other party into an eigenstate of position or momentum by choosing the measurement. There are many criterions for the verification of steering violation, such as the linear steering criterion \cite{PhysRevA.80.032112,PhysRevA.93.020103}, the geometric Bell-like inequalities for steering \cite{PhysRevA.91.032107}, the steering criteria from entropic uncertainty relations \cite{PhysRevLett.106.130402,PhysRevA.87.062103,PhysRevA.98.050104,PhysRevA.98.062111}, and so on.
Quantum steering can be exploited to realize some quantum tasks that classical approach does not work, e.g., quantum information processing \cite{PhysRevLett.107.020401,PhysRevLett.119.110501}, quantum key distribution \cite{PhysRevA.85.010301,PhysRevA.88.052322,Kaur_2020}, and subchannel discrimination \cite{PhysRevLett.114.060404,PhysRevLett.122.130404}.


In particular, a successful and secure quantum information task requires to know how quantum resources are
shared and transformed over many sites.
\textcolor[rgb]{0.00,0.00,0.00}{The question naturally arises of how to enhance one resource by modifying the other, and how much these resources can be converted in practical quantum tasks. }
Recently,
the distribution and transformation of different quantum resources have stimulated a number of studies
\cite{PhysRevLett.115.220501,PhysRevLett.115.020403,PhysRevA.96.032316,PhysRevA.97.052304,PhysRevA.97.042305,Fan_2019,PhysRevLett.125.130401,https://doi.org/10.1002/qute.202100036,PhysRevA.93.062337,PhysRevA.97.042110,PhysRevA.102.052209,PhysRevA.105.022425}.
\textcolor[rgb]{0.00,0.00,0.00}{For example, Svozil\'{\i}k \emph{ et al.} found the conservations between first-order coherence and quantum correlations, including  Bell nonlocality and the degree of entanglement, in two-qubit state case \cite{PhysRevLett.115.220501}. In addition, Kalaga \emph{et al.} investigated the complementary relations among entanglement, coherence, and steering parameter for bipartite subsystems of three-qubit states \cite{kalaga2017quantum,kalaga2022mixedness}.}
\textcolor[rgb]{0.00,0.00,0.00}{The complementary relations among different quantum resources enable one to estimate the degree of one quantum resource for a given degree of another resource, e.g., estimating
entanglement from Bell nonlocality or vice versa \cite{PhysRevA.88.052105}.
Such complementary relations can lead to somehow counterintuitive but sound conclusions that mixed states can be
relatively more entangled \cite{PhysRevA.87.042108} or even more nonclassical \cite{PhysRevA.91.042309}.
}
However, it is worth noting that most of the related
studies are related to the bipartite or three-qubit pure states. Moreover, many of the measures of entanglement chosen of these studies are difficult to analytically calculate for multipartite mixed states.
This raises a significant issue: whether one can find the explicit relations among coherence, entanglement and steering violation for the multipartite mixed states.
In fact, the investigations of the intrinsic relations among various quantum resources in multipartite quantum systems
are especially important for manipulating information transfer and flow in the context of quantum resource theories.

In this paper, we show that complementary relations among different measures of quantum resources exist, including negativity, GBC, first-order coherence, quantum steering, \textcolor[rgb]{0.00,0.00,0.00}{and Bell nonlocality,} for
arbitrary three-qubit states.
The negativity is chosen as the measure of entanglement, because it can be calculated analytically in the multipartite mixed-state scenario.
First of all, we find that the negativity is exactly the same as the GBC for the three-qubit pure states, although the negativity is always less than or equal to the GBC for the mixed states. In addition, the complementary relation between negativity and first-order coherence for tripartite entanglement states are established.
\textcolor[rgb]{0.00,0.00,0.00}{For the three-qubit pure states, we obtain that a single parameter family of state
  $|\psi \rangle_\alpha$ takes the maximum first-order coherence, although another family of state $|\psi \rangle_m$ takes the minimum first-order coherence for a given negativity. Note that state $|\psi \rangle _\alpha$ still take the maximum first-order coherence for the mixed states.}
It is shown that the higher the rank of the density matrix of the state, the closer it is to the origin. Moreover, we study the complementary relation between the negativity and the maximum steering inequality violation. \textcolor[rgb]{0.00,0.00,0.00}{Interestingly, state $|\psi \rangle _m$ takes the maximum steering inequality violation for a given negativity.}
\textcolor[rgb]{0.00,0.00,0.00}{
  Finally, we study the complementary relation between negativity and the maximum Bell-inequality violation in three-qubit quantum systems.}
These relations quantify the intrinsic correlation among these quantum resources
and show how they can be converted from one another.

This paper is organized as follows:
In Sec. \ref{sec2}, we briefly
review some measures of coherence, entanglement, steering inequality violation, \textcolor[rgb]{0.00,0.00,0.00}{and Bell nonlocality.
}
In Sec. \ref{sec3}, we present the close relation between negativity and GBC.
In Sec. \ref{sec4}, we give the complementary relation between the negativity and first-order coherence.
The complementary relation between negativity and the maximum steering inequality violation is studied in Sec. \ref{sec5}. \textcolor[rgb]{0.00,0.00,0.00}{The complementary relation between negativity and the maximum Bell-inequality violation is investigated in Sec. \ref{sec6}.
}
The conclusion is provided in Sec. \ref{sec7}.

\section{PRELIMINARIES\label{sec2}}
Here, we give a brief overview of entanglement, coherence, steering inequality violation, \textcolor[rgb]{0.00,0.00,0.00}{ and Bell nonlocality} to be used in the paper. The measure of entanglement is quantified by the negativity and GBC. The measure of coherence is given by the first-order coherence. We use the three-setting linear steering inequality \textcolor[rgb]{0.00,0.00,0.00}{and Bell-CHSH inequality} as the measures of steering inequality \textcolor[rgb]{0.00,0.00,0.00}{and Bell nonlocality,} respectively.

\subsection{Negativity}
The negativity can be calculated in the same way for pure and mixed
states in arbitrary dimensions. In particular, \textcolor[rgb]{0.00,0.00,0.00}{the tripartite negativity is useful for distillability to a Greenberger-Horne-Zeilinger (GHZ) state in
  quantum computation \cite{PhysRevLett.83.3562}.} For an arbitrary three-qubit state $\rho$, it is defined as \cite{sabin2008classification}
\textcolor[rgb]{0.00,0.00,0.00}{
  \begin{align}
    \mathcal{N}_{A B C}(\rho)=\left(\mathcal{N}_{A|B C} \mathcal{N}_{B|A C} \mathcal{N}_{C|A B}\right)^{\frac{1}{3}},
    \label{Eq.1}
  \end{align}
}
where the bipartite negativity is given by \cite{PhysRevA.65.032314}
\textcolor[rgb]{0.00,0.00,0.00}{
  \begin{align}
    \mathcal{N}_{I|J K}=-2 \sum_{i} N_{i}\left(\rho^{T_I} \right),
    \label{Eq.2}
  \end{align}
}
with ${\rm{I,J,K}} \in \{ A,B,C\} $, ${\rm{I}} \ne {\rm{J}} \ne {\rm{K}}$, and \textcolor[rgb]{0.00,0.00,0.00}{ $N_{i}\left(\rho^{T_I}\right)$} being the negative eigenvalues of the partial transpose \textcolor[rgb]{0.00,0.00,0.00}{$\rho^{T_I}$} of the total state $\rho$ with respect to the subsystem $I$, defined as $\left\langle h_{I}, j_{J K}\left|\textcolor[rgb]{0.00,0.00,0.00}{\rho^{T_I}}\right| k_{I}, l_{J K}\right\rangle=\left\langle k_{I}, j_{J K}|\rho| h_{I}, l_{J K}\right\rangle$.

By the Schmidt decomposition theorem, for any bipartite pure state $|\phi\rangle$ in $d \otimes d'(d \leqslant d')$ quantum system, ${{\cal H}_A} \otimes {{\cal H}_B}$, an alternative form of negativity is written as \cite{PhysRevA.68.062304}
\begin{align}
  \mathcal{N}(|\phi\rangle)=\frac{2}{d-1} \sum_{i<j} \sqrt{\lambda_{i} \lambda_{j}},
  \label{Eq.3}
\end{align}
where $\sqrt{\lambda_{i} }$ and $\sqrt{\lambda_{j}}$ are the Schmidt coefficients, with $\lambda_{i},\lambda_{j}$ being the eigenvalues of the reduced density matrix ${\rho _A}$. For example, if we take $d=2$, then we have
\begin{align}
  \mathcal{N}(|\phi\rangle)=2 \sqrt{\lambda_{1} \lambda_{2}}=2\sqrt {\det {\rho _A}} .
  \label{Eq.4}
\end{align}
Therefore, the negativity of a three-qubit pure state ${\left| \psi  \right\rangle }$ can be rewritten as
\begin{align}
  \begin{split}
    {\cal N}(|\psi \rangle ) = {\left( {\prod\limits_i {2\sqrt {\det {\rho _i}} } } \right) ^{1/3}}
      =2{\left( {\prod\limits_i {\det {\rho _i}} } \right)^{1/6}},
  \end{split}
  \label{Eq.5}
\end{align}
where $i \in \{ A,B,C\} $.

\subsection{GBC}

The GBC is introduced as a genuine multipartite entanglement measure \cite{PhysRevResearch.4.023059}, \textcolor[rgb]{0.00,0.00,0.00}{which should satisfy two conditions: it must be zero for all biseparable states and positive for any nonbiseparable
  state.} The GBC relies on the concept of regularized bipartite concurrence \cite{PhysRevA.71.012318}. The concurrence of a pure bipartite normalized state is given by
\begin{align}
  \mathcal{C}_{A B}(|\psi\rangle)=\sqrt{\frac{d_{\min }}{d_{\min }-1}\left[1-\operatorname{Tr}\left(\rho_{A}^{2}\right)\right]},
  \label{Eq.6}
\end{align}
where $d_{\min }$ denotes the dimension of the smaller subsystem.
For an arbitrary $n$-partite pure state $|\Psi  \rangle$, the
GBC is defined as \cite{PhysRevResearch.4.023059}
\begin{align}
  \mathcal{G}(|\Psi\rangle)=\sqrt[c(\alpha)]{\mathcal{P}(|\Psi\rangle)},
  \label{Eq.7}
\end{align}
\textcolor[rgb]{0.00,0.00,0.00}{where $\alpha  = \left\{ {{\alpha _i}} \right\}$ is the set of all possible bipartitions
$\{ {A_{{\alpha _i}}}|{B_{{\alpha _i}}}\} $ of the $n$ parties, $c(\alpha)$ is the cardinality of $\alpha$
\begin{align}
  c(\alpha) & =\left\{\begin{array}{ll}\vspace{1.5ex}
                        \mathop \sum \limits_{m = 1}^{(n - 1)/2} \left(\!\begin{array}{l}n \\ m\end{array}\!\right),                                                                   & \text { if } n \text { is odd }   \\
                        \mathop \sum \limits_{m = 1}^{(n - 2)/2} \left(\!\begin{array}{c}n \\ m\end{array}\!\right)+\frac{1}{2}\left(\!\begin{array}{c}n / 2 \\ n\end{array}\!\right), & \text { if } n \text { is even. }\end{array}\right.
  \label{Eq.8a}
\end{align}
And ${\cal P}(\left| \Psi  \right\rangle )$ is the product of all bipartite concurrences
\begin{align}
  \mathcal{P}(|\Psi\rangle) & =\prod_{\alpha_{i} \in \alpha} \mathcal{C}_{A_{\alpha_{i}} B_{\alpha_{i}}}(|\Psi\rangle) .
  \label{Eq.8}
\end{align}}
Moreover, the GBC is generalized to mixed states $\rho$ via the convex roof construction
\begin{align}
  \mathcal{G}(\rho)=\inf _{\left\{p_{i},\left|\psi_{i}\right\rangle\right\}}
  \sum_{i} p_{i} \mathcal{G}\left(\left|\psi_{i}\right\rangle\right),
  \label{Eq.12c}
\end{align}
where the infimum is over all feasible decompositions
$\rho  = \sum\limits_i {{p_i}} \left| {{\psi _i}} \right\rangle \langle {\psi _i}|$.

\subsection{First-order coherence}
The first-order coherence is extensively used in the optical systems for the measure of coherence. Its quantification is independent of the selection of the reference basis.
For an arbitrary three-qubit state $\rho_{ABC}$, the first-order coherence of its subsystems $A$, $B$ and $C$ are defined, in terms of the purity \cite{mandel_wolf_1995}, as
\begin{align}
  {\cal D}\left(\rho_{i}\right)=\sqrt{2 \operatorname{Tr}\left(\rho_{i}^{2}\right)-1},
  \label{Eq.D1}
\end{align}
where $i \in \{ A,B,C\} $.
When all subsystems can be considered independently, the first-order coherence of state $\rho_{ABC}$ is given by \cite{PhysRevLett.115.220501}
\begin{align}
  {\cal D}\left(\rho_{A B C}\right)=\sqrt{\frac{{\cal D}\left(\rho_{A}\right)^{2}+{\cal D}\left(\rho_{B}\right)^{2}+{\cal D}\left(\rho_{C}\right)^{2}}{3}},
  \label{Eq.Dabc}
\end{align}
where $0 \leqslant {\cal D}({\rho _{ABC}}) \leqslant 1$.

\subsection{The three-setting linear-steering inequality violation}
Quantum steering is considered as a subset of entanglement and a superset
  of Bell nonlocality \cite{PhysRevA.92.032107}. From the local hidden states model, some steering inequalities are derived to indicate steering phenomenon by the violation of them. As an example, \textcolor[rgb]{0.00,0.00,0.00}{the linear-steering inequality is formulated by Cavalcanti \emph{et al.}
to verify whether a bipartite state is steerable when Alice and
Bob are both allowed to operate $n$ dichotomic
measurements on their own subsystems \cite{PhysRevA.80.032112}} :
\begin{align}
  F_{n}\left(\rho_{A B}, \mu\right)=\frac{1}{\sqrt{n}}\left|\sum_{k=1}^{n}\left\langle A_{k} \otimes B_{k}\right\rangle\right| \leqslant 1,
  \label{Eq.Fn}
\end{align}
where $A_{k}=\hat{a}_{k} \cdot \vec{\sigma}$ and $B_{k}=\hat{b}_{k} \cdot \vec{\sigma}$,
with $\vec{\sigma}=\left(\sigma_{1}, \sigma_{2}, \sigma_{3}\right)$ being
the Pauli matrices; $\hat{a}_{k}, \hat{b}_{k} \in \mathbb{R}^{3}$
are unit and orthonormal vectors; $\,\left\langle A_{k} \otimes B_{k}\right\rangle=\operatorname{Tr}\left(\rho_{A B}\left(A_{k} \otimes B_{k}\right)\right)$;
and $\mu=\left\{\hat{a}_{1}, \hat{a}_{2}, \ldots, \hat{a}_{n}, \hat{b}_{1}, \hat{b}_{2}, \ldots, \hat{b}_{n}\right\}$
is the set of measurement directions.

\textcolor[rgb]{0.00,0.00,0.00}{
In general, an arbitrary two-qubit state can be denoted by the
Hilbert-Schmidt representation
\begin{align}
  \rho_{A B}\!=\!\frac{1}{4}\!\left[I_{2} \otimes I_{2}\!+\!\vec{a} \cdot \vec{\sigma} \otimes I_{2}\!
  +\!I_{2} \otimes \vec{b} \cdot \vec{\sigma}\!+\!\sum_{i, j} t_{i j} \sigma_{i} \otimes \sigma_{j}\right],
  \label{Eq.rhoAB}
\end{align}
where $\vec{a}$ and $\vec{b}$ are the local bloch vectors, and ${T_{AB}} = [{t_{ij}}]$ is the correlation matrix. The components ${t_{ij}}$ are given by ${t_{ij}} = {\mathop{\rm Tr}\nolimits} \left( {{\rho _{AB}}\left( {{\sigma _i} \otimes {\sigma _j}} \right)} \right)$.
        For the three measurement settings corresponding $n=3$ of Eq. (\ref{Eq.Fn}), state $\rho _{AB}$ is $F_3$ steerable if \cite{PhysRevA.93.020103,PhysRevA.102.052209}
    }
    \textcolor[rgb]{0.00,0.00,0.00}{
  \begin{align}
    {\cal S}_{A B}=\operatorname{Tr}\left(T_{A B}^{T} T_{A B}\right)-1>0,
    \label{Eq.Sab}
  \end{align}
  }
  where the superscript $T$ represents the transpose of the correlation matrix $T_{AB}$.
  \textcolor[rgb]{0.00,0.00,0.00}{It can be shown that this steering inequality is a two-way steering criterion due to its invariance under qubit permutations.}
  Among the three bipartite reduced
  states of a three-qubit state ${\rho _{ABC}}$, the maximum steering inequality
  violation is given by \cite{PhysRevA.102.052209}
  \begin{align}
    {\cal S}\left(\rho_{A B C}\right)=\max \left\{{\cal S}_{A B}, {\cal S}_{A C},  {\cal S}_{B C}\right\}.
    \label{Eq.Sabc}
  \end{align}

  \textcolor[rgb]{0.00,0.00,0.00}{
  \subsection{Bell-inequality violation}
  In 1995, Horodecki \emph{et al.} presented the necessary and sufficient condition for violating the Bell-CHSH inequality \cite{HORODECKI1995340}. For an arbitary two-qubit state $\rho_{AB}$, the maximum Bell-CHSH value ${\cal B}_{AB}^{'}$ is given by
  \begin{align}
    {\cal B}_{AB}^{'} = 2\sqrt {M_{AB}},
    \label{Eq.16h}
  \end{align}
  where $M_{AB} = {m_1} + {m_2}$, with ${m_1}$ and ${m_2}$ being the largest two
  eigenvalues of $T_{AB} ^T{T_{AB} }$, in which $T_{AB}$ is the correlation matrix.
$M_{AB} > 1$ implies the violation of the Bell-CHSH inequality. In this case, the Bell-inequality violation (i.e. the Bell-CHSH inequality violation) ${\cal B}_{AB}$ is defined
  as \cite{Sadhukhan_2015}
  \begin{align}
    {\cal B}_{AB}=\max \left\{0, M_{AB}-1\right\}.
    \label{Eq.17h}
  \end{align}
  Among the three pairwise reduced states of a three-qubit
  state ${\rho _{ABC}}$, the maximum Bell-inequality violation is obtained as \cite{PhysRevA.94.052126}
  \begin{align}
    \mathcal{B}({\rho _{ABC}})=\max \left\{\mathcal{B}_{A B}, \mathcal{B}_{B C}, \mathcal{B}_{A C}\right\},
    \label{Eq.16a}
  \end{align}
  where only one of ${\cal B}_{AB}$, ${\cal B}_{AC}$ and ${\cal B}_{BC}$ is nonzero \cite{toner2006monogamy}.
  }

  \subsection{Two useful boundary states}

  In order to express the complementary relations of the above quantum resources for the arbitrary three-qubit states in a more explicit manner, here we introduce two boundary states with a single parameter. The first one is the generalized GHZ state, which can exhibit
  maximum first-order coherence value for a fixed amount of negativity,
  \begin{align}
    {\left| \psi  \right\rangle _\alpha } = \cos \alpha \left| {i,j,k} \right\rangle  + \sin \alpha {\rm{ }}\left| {\bar i,\bar j,\bar k} \right\rangle ,
    \label{Eq.21a}
  \end{align}
  where $i,j,k \in \{ 0,1\} $ and the overbar means taking the opposite value.
  In the following, we take states with $i=j=k=0$ as an example in the calculation:
  \begin{align}
    {\left| \psi  \right\rangle _\alpha } = \cos \alpha \left| {000} \right\rangle  + \sin \alpha \left| {111} \right\rangle .
    \label{Eq.22a}
  \end{align}

  The second boundary state is a single parameter family of three-qubit pure state
  \begin{align}
    |\psi\rangle_{m}=\frac{|000\rangle+m(|010\rangle+|101\rangle)+|111\rangle}{\sqrt{2+2 m^{2}}},
    \label{Eq.23a}
  \end{align}
  where $m \in [0,1]$. Note that the state is a GHZ-class state when $m \in [0,1)$, and it is a $W$-class state when $m = 1$.

  \section{negativity versus the GBC\label{sec3}}
   {\it Theorem 1.\,}
  For the three-qubit pure states, the negativity is exactly \textcolor[rgb]{0.00,0.00,0.00}{equivalent} to the GBC. However, for a three-qubit mixed state $\rho$, the negativity is always less than or equal to the GBC,
  \begin{align}
    {\cal N}(\rho ) \leqslant {\cal G}(\rho ).
    \label{Eq.22c}
  \end{align}

  \begin{proof}
    For a three-qubit pure state $|\psi \rangle$, the GBC is given by
    \begin{align}
      \begin{split}
        {\cal G}(|\psi \rangle ) &= {\left\{ {\prod\limits_i {\sqrt {2\left[ {1 - \operatorname{Tr}(\rho _i^2)} \right]} } } \right\}^{1/3}},
        \label{Eq.9}
      \end{split}
    \end{align}
    where $i \in \{ A,B,C\} $. Due to the trace condition of the reduced density matrices, ${\lambda _1} + {\lambda _2} = 1$, we can obtain that
    \begin{align}
      \begin{split}
        \sqrt {2\left[ {1 - \operatorname{Tr}\left( {\rho _A^2} \right)} \right]}  &= \sqrt {2\left[ {1 - \left( {\lambda _1^2 + \lambda _2^2} \right)} \right]}\\
        & = 2\sqrt {{\lambda _1}{\lambda _2}}=2\sqrt {\det {\rho _A}} .
        \label{Eq.10}
      \end{split}
    \end{align}
    Similarly,
    \begin{align}
      \begin{split}
        \sqrt {2\left[ {1 - \operatorname{Tr}\left( {\rho _B^2} \right)} \right]}  = 2\sqrt {\det {\rho _B}} \\
        \sqrt {2\left[ {1 - \operatorname{Tr}\left( {\rho _C^2} \right)} \right]}  = 2\sqrt {\det {\rho _C}}.
        \label{Eq.11}
      \end{split}
    \end{align}
    Substituting Eqs. (\ref{Eq.10}) and (\ref{Eq.11}) into Eq. (\ref{Eq.5}), we have
    \begin{align}
      {\cal N}(|\psi \rangle )={\cal G}(|\psi \rangle ).
      \label{Eq.20a}
    \end{align}
    For a three-qubit mixed state $\rho$, since the bipartite negativity is a convex function \cite{PhysRevA.65.032314}, the tripartite negativity, as the geometric mean of three bipartite negativities, is also convex function  \cite{niculescu2006convex}. In addition, the GBC is defined as the minimum decomposition $\sum\nolimits_j {{p_j}\left| {{\psi _j}} \right\rangle \langle {\psi _j}|} $ over all feasible decompositions. Thus, we obtain the following relation
    \begin{align}
      {\cal N}(\rho ) \leqslant \sum\limits_j {{p_j}{\cal N}({\psi _j})}  = \sum\limits_j {{p_j}{\cal G}({\psi _j})}  = {\cal G}(\rho ).
      \label{Eq.27c}
    \end{align}
  \end{proof}

  \section{negativity versus first-order coherence \label{sec4} }

   {\it Theorem 2.\,}
  If a three-qubit pure state $|\psi \rangle$ has
  the same value of negativity with states $|\psi \rangle _{\alpha}$ and $|\psi \rangle _m$, the first-order coherence
  of these three states satisfies the ordering ${\cal D}{(|\psi \rangle _m}) \leqslant {\cal D}(|\psi \rangle ) \leqslant {\cal D}{(|\psi \rangle _{\alpha}})$.
  The complementary relation of negativity and first-order coherence is expressed as
  \begin{align}
     & {\cal N}{\left( {\left| \psi  \right\rangle } \right)^2} + {\cal D}{\left( {\left| \psi  \right\rangle } \right)^2}\leqslant 1,\nonumber \\
     & {\cal N}{\left( {\left| \psi  \right\rangle } \right)^6} + 3{\cal D}{\left( {\left| \psi  \right\rangle } \right)^2} \geqslant 1.
    \label{Eq.30}
  \end{align}
  \textcolor[rgb]{0.00,0.00,0.00}{Note that} the first inequality is still valid for the three-qubit mixed states.

  \begin{proof}
    \textcolor[rgb]{0.00,0.00,0.00}{We will prove this theorem for the pure states first, and then extend it to the case of the mixed states.}

    Combining Eqs. (\ref{Eq.D1}) and (\ref{Eq.Dabc}), we have
    \begin{align}
      {\cal D}\left( \rho \right)^2 = \frac{2}{3}\sum\limits_i {\operatorname{Tr}(\rho _i^2)}  - 1,
      \label{Eq.31}
    \end{align}
    where $i \in \{ A,B,C\} $. Then we can construct a equation
    \begin{align}
      \sum\limits_i {\left[ {1 - \operatorname{Tr}(\rho _i^2)} \right]}  + \sum\limits_i {\operatorname{Tr}(\rho _i^2)}  = 3.
      \label{Eq.32}
    \end{align}
    and using the arithmetic-geometric mean value inequality, we have
    \begin{align}
      \begin{split}
        \frac{1}{3}\sum\limits_i {2\left[ {1 - \operatorname{Tr}(\rho _i^2)} \right]} \geqslant {\left\{ {\prod\limits_i {2\left[ {1 - \operatorname{Tr}(\rho _i^2)} \right]} } \right\}^{1/3}} .
      \end{split}
      \label{Eq.A2}
    \end{align}
    By adding a term $2\sum\limits_i {\operatorname{Tr}(\rho _i^2)}/3  - 1$ to both sides of the above equation, we have
    \begin{align}
      {\left\{ {\prod\limits_i {\sqrt {2\left[ {1 - \operatorname{Tr}(\rho _i^2)} \right]} } } \right\}^{2/3}} + \frac{2}{3}\sum\limits_i {\operatorname{Tr}(\rho _i^2)}  - 1 \leqslant 1.
      \label{Eq.33}
    \end{align}
    For a three-qubit pure state $\left| \psi  \right\rangle$, substituting Eqs. (\ref{Eq.9}) and
    (\ref{Eq.31}) into Eq. (\ref{Eq.33}), and replace ${\cal G}\left( \left| \psi  \right\rangle \right)$ with ${\cal N}\left( \left| \psi  \right\rangle \right)$, we get
    \begin{align}
      {\cal N}{\left( {\left| \psi  \right\rangle } \right)^2} + {\cal D}{\left( {\left| \psi  \right\rangle } \right)^2} \leqslant 1.
      \label{Eq.34}
    \end{align}
    To verify the second inequality in Eq. (\ref{Eq.30}), we can construct a function $H(u,v,w)$ as
    \begin{align}
      H(u,v,w) = {2^4}uvw - u - v - w + \frac{1}{2},
      \label{Eq.A4}
    \end{align}
    where $u,v,w \in [0,\frac{1}{4}]$. It can be found that
    \begin{align}
      \begin{split}
        \frac{{\partial H}}{{\partial u}} = {2^4}vw - 1 \leqslant 0\\
        \frac{{\partial H}}{{\partial v}} = {2^4}uw - 1 \leqslant 0\\
        \frac{{\partial H}}{{\partial w}} = {2^4}uv - 1 \leqslant 0.
      \end{split}
      \label{Eq.A5}
    \end{align}
    Then the minimum of the function $H$ can be calculated as
    \begin{align}
      H(\frac{1}{4},\frac{1}{4},\frac{1}{4}) = 0.
      \label{Eq.A6}
    \end{align}
    On the other hand, the trace conditions of the reduced density matrices $\rho _A$, $\rho _B$ and $\rho _C$ are
    \begin{align}
      {\lambda _1} + {\lambda _2} = 1,\quad {\lambda _3} + {\lambda _4} = 1,\quad {\lambda _5} + {\lambda _6} = 1,
      \label{Eq.A7}
    \end{align}
    where $\lambda _1$ and $\lambda _2$, $\lambda _3$ and $\lambda _4$, and $\lambda _5$ and $\lambda _6$ are the eigenvalues of $\rho _A$, $\rho _B$, and $\rho _C$, respectively. We can see that
    \begin{align}
      0 \leqslant {\lambda _1}{\lambda _2} = {\lambda _1}(1 - {\lambda _1}) \leqslant \frac{1}{4}.
      \label{Eq.A8}
    \end{align}
    Similarly, we have
    \begin{align}
      0 \leqslant {\lambda _3}{\lambda _4} \leqslant \frac{1}{4},\quad 0 \leqslant {\lambda _5}{\lambda _6} \leqslant \frac{1}{4},
      \label{Eq.A9}
    \end{align}
    Let $u={\lambda _1}{\lambda _2}$, $v={\lambda _3}{\lambda _4}$, and $w={\lambda _5}{\lambda _6}$, and substituting them into Eq. (\ref{Eq.A4}), we obtain
    \begin{align}
      {2^4}\prod\limits_{\mu=1}^6 {{\lambda _\mu}}  - {\lambda _1}{\lambda _2} - {\lambda _3}{\lambda _4} - {\lambda _5}{\lambda _6} + \frac{1}{2} \geqslant 0.
      \label{Eq.35}
    \end{align}
    Then, by using the trace conditions of the reduced density matrices, we get
    \begin{align}
      {2^6}\prod\limits_{\mu=1}^6 {{\lambda _\mu }}  + 2\left( {\lambda _1^2 + \lambda _2^2 + \lambda _3^2 + \lambda _4^2 + \lambda _5^2 + \lambda _6^2} \right) - 3 \geqslant 1.
      \label{Eq.A13}
    \end{align}
    Finally, we can find that
    \begin{align}
      {\left[ {2{{\left( {\prod\limits_i {\det {\rho _i}} } \right)}^{1/6}}} \right]^6} + 3\left[ {\frac{2}{3}\sum\limits_i {\operatorname{Tr}(\rho _i^2)}  - 1} \right] \geqslant 1.
      \label{Eq.36}
    \end{align}
    Substituting Eqs. (\ref{Eq.5}) and
    (\ref{Eq.31}) into Eq. (\ref{Eq.36}), we can obtain that
    \begin{align}
      {\cal N}{\left( {\left| \psi  \right\rangle } \right)^6} + 3{\cal D}{\left( {\left| \psi  \right\rangle } \right)^2} \geqslant 1.
      \label{Eq.37}
    \end{align}
    On the other hand, the negativity and first-order coherence of the boundary states $|\psi \rangle _{\alpha}$ and $|\psi \rangle _{m}$, from Eqs. (\ref{Eq.5}) and (\ref{Eq.Dabc}), are given by
    \begin{align}
      {\cal N}{(|\psi \rangle _\alpha }) = \left| {\sin 2\alpha } \right|,
      \label{Eq.38}
    \end{align}
    \begin{align}
      {\cal D}{(|\psi \rangle _\alpha }) = \left| {\cos 2\alpha } \right| ,
      \label{Eq.39}
    \end{align}
    \begin{align}
      {\cal N}{(|\psi \rangle _m })  = {\left( {\frac{{1 - {m^2}}}{{1 + {m^2}}}} \right)^{1/3}},
      \label{Eq.40}
    \end{align}
    \begin{align}
      {\cal D}{(|\psi \rangle _m}) = \frac{{2m}}{{\sqrt 3 (1 + {m^2})}}.
      \label{Eq.41}
    \end{align}
    We can find that
    \begin{align}
      {\cal N}{\left( {\left| \psi  \right\rangle _\alpha } \right)^2} + {\cal D}{\left( {\left| \psi  \right\rangle _\alpha} \right)^2} = 1,
      \label{Eq.42}
    \end{align}
    \begin{align}
      {\cal N}{\left( {\left| \psi  \right\rangle _m} \right)^6} + 3{\cal D}{\left( {\left| \psi  \right\rangle _m} \right)^2} = 1,
      \label{Eq.43}
    \end{align}
    which imply that states $|\psi \rangle _{\alpha}$ and $|\psi \rangle _{m}$ are the upper and lower boundary states, respectively.

    \textcolor[rgb]{0.00,0.00,0.00}{
    It can be found that the first-order coherence of subsystems of a three-qubit mixed state $\rho _{ABC}$ are convex functions.
    The first-order coherence of state $\rho _{ABC}$, as the vector composition of ${\cal D}\left(\rho_{i}\right)$ and $h({x_1},{x_2},{x_3}) = {\left[ {\left( {x_1^2 + x_2^2 + x_3^2} \right)/3} \right]^{1/2}}$, is also a convex function \cite{boyd2004convex}. On the other hand, Eq. (\ref{Eq.34}) can be rewritten as
    \begin{align}
      {\cal D}\left( {\left| \psi  \right\rangle } \right) \leqslant \sqrt {1 - {\cal N}{{\left( {\left| \psi  \right\rangle } \right)}^2}}.
      \label{Eq.43b}
    \end{align}
    Let $U\left[ {{\cal N}\left( {\left| \psi  \right\rangle } \right)} \right] = \sqrt {1 - {\cal N}{{\left( {\left| \psi  \right\rangle } \right)}^2}} $, then we can see that the function $U$ is concave function in regard to ${\cal N}{\left( {\left| \psi  \right\rangle } \right)}$. Using the convexity of negativity, we have
    \begin{align}
      \begin{split}
        {\cal D}\left( \rho  \right) \leqslant \sum\limits_i {{p_i}{\cal D}({\psi _i})} & \leqslant \sum\limits_i {{p_i}\sqrt {1 - {\cal N}{{\left( {{\psi _i}} \right)}^2}} } \\
        &\leqslant \sqrt {1 - {{\left[ {\sum\limits_i {{p_i}} {\cal N}\left( {{\psi _i}} \right)} \right]}^2}} \\
        &\leqslant \sqrt {1 - {\cal N}{{\left( \rho  \right)}^2}},
      \end{split}
      \label{Eq.44}
    \end{align}
    i.e.,
    \begin{align}
      {\cal N}{\left( \rho  \right)^2} + {\cal D}{\left( \rho  \right)^2} \leqslant 1.
      \label{Eq.45c}
    \end{align}
    }
  \end{proof}

  \begin{figure}
    \centering
    \includegraphics[width=8.6cm]{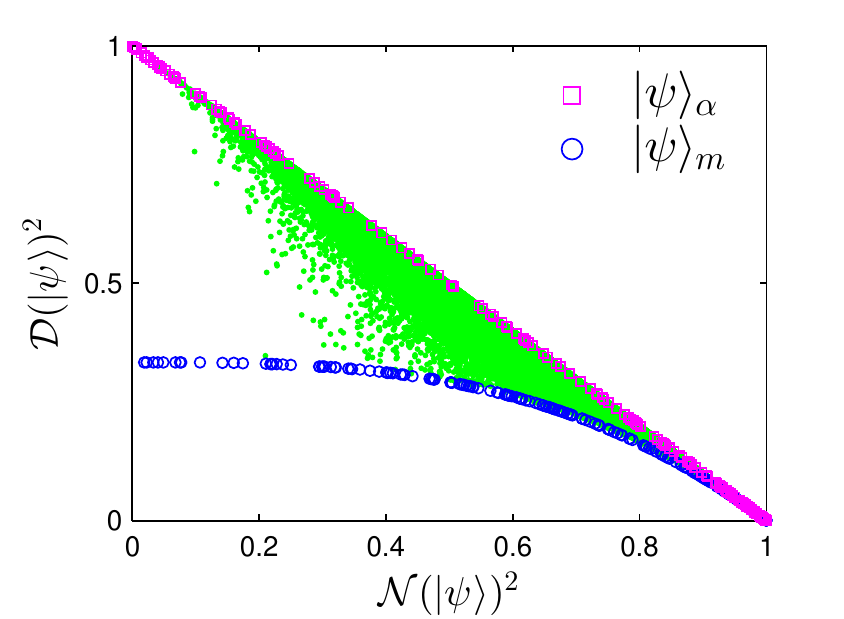}
    \caption{(Color online)  Complementary relation between the negativity ${\cal N}(|\psi \rangle )$ and the first-order coherence ${\cal D}(|\psi \rangle )$ for ${10^5}$ Haar randomly generated three-qubit pure states. \textcolor[rgb]{0.00,0.00,0.00}{The magenta squares are located at the upper boundary with the sate $|\psi \rangle _\alpha$, and state $|\psi \rangle _m$ represented by blue circles lies at the lower boundary.} Both axes are dimensionless.}
    \label{f1}
  \end{figure}

  In Fig. \ref{f1}, we plot how the square of first-order coherence changes with respect to the square of negativity for ${10^5}$ Haar randomly generated three-qubit pure states \cite{bengtsson_zyczkowski_2017,*Zyczkowski_1994}. \textcolor[rgb]{0.00,0.00,0.00}{The magenta squares donating state $|\psi \rangle _\alpha$ are located at the upper boundary, which satisfies the relation between negativity and first-order coherence in Eq. (\ref{Eq.42}). The blue circles at the lower boundary show that the two quantum resources of state $|\psi \rangle _m$ fulfill the
    relation in Eq. (\ref{Eq.43}).} The ${10^5}$ Haar randomly generated three-qubit pure states are included in the range constrained  by states $|\psi \rangle _\alpha$ and $|\psi \rangle _m$, meaning that their negativity and first-order coherence obey the inequalities in Eq. (\ref{Eq.30}). Moreover, we find that the first-order coherence increases (decreases) with the decreases (increases) of the negativity, shows a complementary.

\begin{figure}
    \centering
    \includegraphics[width=8.6cm]{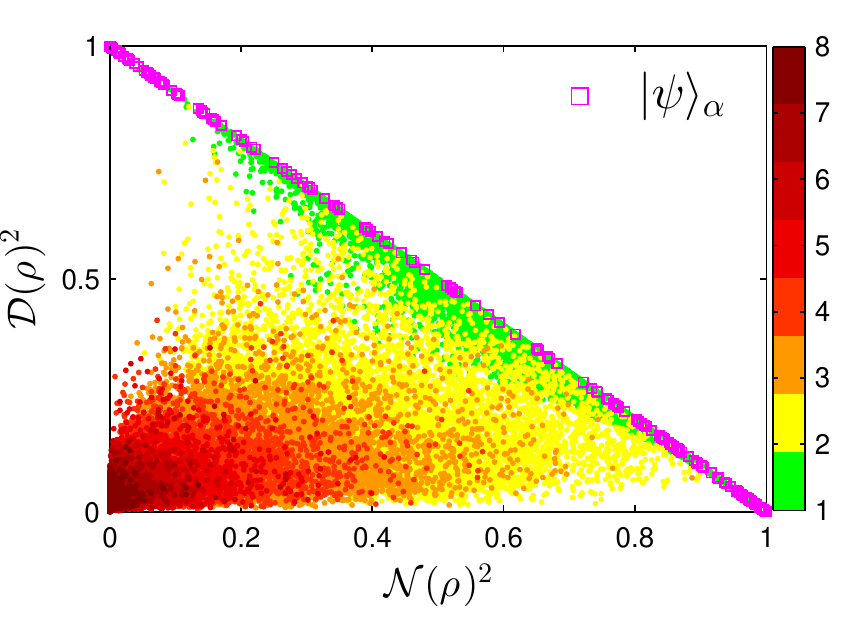}
    \caption{(Color online)  Complementary relation between the negativity ${\cal N}(\rho)$ and the first-order coherence ${\cal D}(\rho )$ for ${10^5}$ Haar randomly generated three-qubit mixed states. \textcolor[rgb]{0.00,0.00,0.00}{The magenta squares are located at the upper boundary with sate $|\psi \rangle _\alpha$.} The numbers on the color bar represent different ranks of the density matrixes of the random states. Both axes are dimensionless.}
    \label{f2}
  \end{figure}

  Figure \ref{f2} plots the relation between negativity and first-order coherence for ${10^5}$ Haar randomly generated three-qubit mixed states. \textcolor[rgb]{0.00,0.00,0.00}{The magenta squares are still located at the upper boundary with the sate $|\psi \rangle _\alpha$.} The ${10^5}$ Haar randomly generated three-qubit mixed states are under the boundary line, which means that their negativity and first-order coherence satisfy the inequality in Eq. (\ref{Eq.45c}). In addition, we can see that the higher the rank of the density matrix of the random state, the closer it is to the origin. Also, it shows that a complementary relation between the negativity and first-order coherence for arbitrary three-qubit states exists.

  \section{negativity VERSUS MAXIMUM STEERING INEQUALITY VIOLATION \label{sec5}}
   {\it Theorem 3.\,}
  If an arbitrary three-qubit state $\rho$ has
  the same value of negativity with state $|\psi \rangle _m$, the maximum steering inequality violations
  of these two states satisfy the ordering $  {\cal S}(\rho ) \leqslant {\cal S}{(|\psi \rangle _m})$.
  The complementary relation of the negativity and the maximum steering inequality violation is given by
  \textcolor[rgb]{0.00,0.00,0.00}{
  \begin{align}
    2{\cal N}{\left( {\rho } \right)^6} + {\cal S}\left( {\rho } \right) \leqslant 2.
    \label{Eq.44a}
  \end{align}
  }

  \begin{proof}
    \textcolor[rgb]{0.00,0.00,0.00}{An arbitrary three-qubit state ${\rho _{ABC}}$ can be written as
      \begin{align}
        \begin{split}
          \rho_{A B C}=& \frac{1}{8} \Bigg[ \mathbb{I} \otimes \mathbb{I} \otimes \mathbb{I}+\vec{A} \cdot \vec{\sigma} \otimes \mathbb{I} \otimes \mathbb{I}+\mathbb{I} \otimes \vec{B} \cdot \vec{\sigma} \otimes \mathbb{I}\\
            &+\mathbb{I} \otimes \mathbb{I} \otimes \vec{C} \cdot \vec{\sigma}+\sum_{i j} t_{i j}^{A B} \sigma_{i} \otimes \sigma_{j} \otimes \mathbb{I} \\
            &+\sum_{i k} t_{i k}^{A C} \sigma_{i} \otimes \mathbb{I} \otimes \sigma_{k}+\sum_{j k} t_{j k}^{B C} \mathbb{I} \otimes \sigma_{j} \otimes \sigma_{k} \\
            &\left.+\sum_{i j k} t_{i j k}^{A B C} \sigma_{i} \otimes \sigma_{j} \otimes \sigma_{k} \right].
        \end{split}
        \label{Eq.45a}
      \end{align}
      The purities of the the reduced density matrices $\rho_{A}$ and $\rho_{B C}$ are
      \textcolor[rgb]{0.00,0.00,0.00}{
      \begin{align}
        \operatorname{Tr}\left(\rho_{A}^{2}\right)\!\!=\!\!\frac{1+\vec{A}^{2}}{2}, \quad \operatorname{Tr}\left(\rho_{B C}^{2}\right)\!\!=\!\!\frac{1}{4}\left(2+\vec{B}^{2}+\vec{C}^{2}+{\cal S}_{B C}\right).
        \label{Eq.46a}
      \end{align}
      }
      Similarly, we obtain
      \textcolor[rgb]{0.00,0.00,0.00}{
      \begin{align}
        \operatorname{Tr}\left(\rho_{B}^{2}\right)\!\!=\!\!\frac{1+\vec{B}^{2}}{2}, \quad \operatorname{Tr}\left(\rho_{A C}^{2}\right)\!\!=\!\!\frac{1}{4}\left(2+\vec{A}^{2}+\vec{C}^{2}+{\cal S}_{A C}\right)\nonumber,
      \end{align}
      \begin{align}
        \operatorname{Tr}\left(\rho_{C}^{2}\right)\!\!=\!\!\frac{1+\vec{C}^{2}}{2}, \quad \operatorname{Tr}\left(\rho_{A B}^{2}\right)\!\!=\!\!\frac{1}{4}\left(2+\vec{A}^{2}+\vec{B}^{2}+{\cal S}_{A B}\right).
        \label{Eq.47a}
      \end{align}
      }
      \textcolor[rgb]{0.00,0.00,0.00}{In the following, we will give the proof for pure states first, and then extend the theorem to mixed states.}
      If $\rho_{ABC}$ is a pure state with $\rho_{ABC}=|\psi \rangle\langle\psi|$, based on the Schmidt decomposition, we have $\operatorname{Tr}(\rho _i^2) = \operatorname{Tr}(\rho _{jk}^2)$
      for i $\ne j \ne k,\,\,i,j,k \in \{ A,B,C\} $. By Eqs. (\ref{Eq.46a}) and (\ref{Eq.47a}),
      the linear steering  inequality violation of the bipartite reduced states of $\rho_{AB}$, ${\cal S}_{AB}$, can be written as a function of purities of subsystems of state $|\psi \rangle$
      \textcolor[rgb]{0.00,0.00,0.00}{
      \begin{align}
        {{\cal S}_{AB}} = 2[2\operatorname{Tr}(\rho _C^2) - \operatorname{Tr}(\rho _A^2) - \operatorname{Tr}(\rho _B^2)] .
        \label{Eq.48a}
      \end{align}}
      }
    Assuming that ${\cal S}(|\psi \rangle)  = {{\cal S}_{AB}}$, then let us construct a function with the form
    \begin{align}
      R(u,v,w) = {2^6}uvw + 2u + 2v - 4w,
      \label{Eq.B1}
    \end{align}
    where $u,v,w \in [0,\frac{1}{4}]$. We can show that
    \begin{align}
      \begin{split}
        \frac{{\partial R}}{{\partial u}} = {2^6}vw + 2 \geqslant 0\\
        \frac{{\partial R}}{{\partial v}} = {2^6}uw + 2 \geqslant 0\\
        \frac{{\partial R}}{{\partial w}} = {2^6}uv - 4 \leqslant 0.
      \end{split}
      \label{Eq.B2}
    \end{align}
    Thus the maximum of the function $R$ is given by
    \begin{align}
      R(\frac{1}{4},\frac{1}{4},0) = 1.
      \label{Eq.B3}
    \end{align}
    Substituting relations $u={\lambda _1}{\lambda _2}$, $v={\lambda _3}{\lambda _4}$ and $w={\lambda _5}{\lambda _6}$ into $R(u,v,w)$ in Eq. (\ref{Eq.B1}), we can obtain
    an inequality with respect to the eigenvalues of the reduced density matrices as
    \begin{align}
      {2^6}\prod\limits_{\mu=1}^6 {{\lambda _\mu}}  + 2{\lambda _1}{\lambda _2} + 2{\lambda _3}{\lambda _4} - 4{\lambda _5}{\lambda _6} \leqslant 1.
      \label{Eq.49a}
    \end{align}
    The above inequality can be rewritten as
    \begin{align}
      {2^6}\prod\limits_{\mu  = 1}^6 {{\lambda _\mu }}\!  +\! 2\left( {1 - 2{\lambda _5}{\lambda _6}} \right) \!-\! \left( {1\! -\! 2{\lambda _1}{\lambda _2}} \right) \!-\! \left( {1 - 2{\lambda _3}{\lambda _4}} \right) \leqslant 1.
      \label{Eq.B6}
    \end{align}
    By using the trace conditions of the reduced density matrices, we have
    \begin{align}
      {2^6}\prod\limits_{\mu  = 1}^6 {{\lambda _\mu }}  + 2\left( {\lambda _5^2 + \lambda _6^2} \right) - \left( {\lambda _1^2 + \lambda _2^2} \right) - \left( {\lambda _3^2 + \lambda _4^2} \right) \leqslant 1.
      \label{Eq.B7}
    \end{align}
    Finally, we can get
    \textcolor[rgb]{0.00,0.00,0.00}{
    \begin{align}
      \begin{split}
        2{\left[ {2{{\left( {\prod\limits_i {\det {\rho _i}} } \right)}^{1/6}}} \right]^6}&+ 2[2\operatorname{Tr}(\rho _C^2)\\
        &- \operatorname{Tr}(\rho _A^2) - \operatorname{Tr}(\rho _B^2)]  \leqslant 2.
        \label{Eq.50a}
      \end{split}
    \end{align}
    }
    Substituting Eqs. (\ref{Eq.5}) and (\ref{Eq.48a}) into Eq. (\ref{Eq.50a}), we have
    \textcolor[rgb]{0.00,0.00,0.00}{
    \begin{align}
      2{\cal N}{\left( {\left| \psi  \right\rangle } \right)^6} + {\cal S}\left( {\left| \psi  \right\rangle } \right) \leqslant 2.
      \label{Eq.51a}
    \end{align}
    }
    The complementary relation also holds if ${\cal S}(|\psi \rangle)  = {{\cal S}_{AC}}$ or ${\cal S}(|\psi \rangle)  = {{\cal S}_{BC}}$.

    Moreover, the maximum steering inequality violation of the boundary state $|\psi \rangle _{m}$, from Eq. (\ref{Eq.Sabc}), can be calculated as
    \textcolor[rgb]{0.00,0.00,0.00}{
    \begin{align}
      {\cal S}{(|\psi \rangle _m}) = \frac{{ 8m^2}}{{{(1 + {m^2})^2}}}.
      \label{Eq.52a}
    \end{align}
    }
    Together with Eq. (\ref{Eq.40}), we can obtain
    \textcolor[rgb]{0.00,0.00,0.00}{
    \begin{align}
      2{\cal N}{\left( {\left| \psi  \right\rangle _m } \right)^6} + {\cal S}\left( {\left| \psi  \right\rangle _m } \right) = 2,
      \label{Eq.53a}
    \end{align}
    }
    which imply that state $|\psi \rangle _{m}$ is the upper boundary states.

    On the other hand, Eq. (\ref{Eq.51a}) can be rewritten as
    \textcolor[rgb]{0.00,0.00,0.00}{
    \begin{align}
      {\cal S}\left( {\left| \psi  \right\rangle } \right) \leqslant 2 [1- {\cal N}{\left( {\left| \psi  \right\rangle } \right)^6}].
      \label{Eq.54a}
    \end{align}
    }
    Let \textcolor[rgb]{0.00,0.00,0.00}{$L\left[ {{\cal N}\left( {\left| \psi  \right\rangle } \right)} \right] = 2 [1- {\cal N}{\left( {\left| \psi  \right\rangle } \right)^6}]$}, we can show that $L$ is a concave function with respect to ${\cal N}\left( {\left| \psi  \right\rangle } \right)$.
    \textcolor[rgb]{0.00,0.00,0.00}{
      If $\rho_{ABC}$ is a mixed state, both its negativity and the maximum steering inequality violation are convex functions \cite{PhysRevA.102.052209}. Similar to the derivation in Eq. (\ref{Eq.44}), we can obtain the complementary relation between the negativity and the maximum steering inequality violation for the three-qubit mixed states as
      \textcolor[rgb]{0.00,0.00,0.00}{
      \begin{align}
        2{\cal N}{\left( \rho  \right)^6} + {\cal S}\left( \rho  \right) \leqslant 2.
        \label{Eq.57a}
      \end{align}
      }
    }
  \end{proof}

  \begin{figure}
    \centering
    \includegraphics[width=8.6cm]{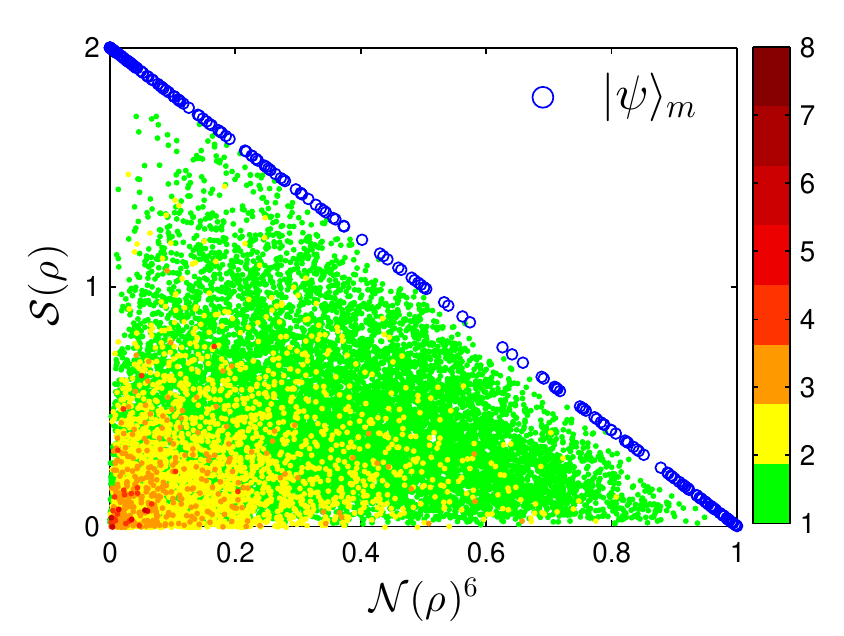}
    \caption{(Color online)  Complementary relation between the negativity ${\cal N}(\rho )$ and the maximum steering inequality violation ${\cal S}(\rho )$ for ${10^5}$ Haar randomly generated three-qubit mixed states of ${\cal S}(\rho ) \geqslant 0$. \textcolor[rgb]{0.00,0.00,0.00}{The blue circles are located at the upper boundary with state $|\psi \rangle _m$.} The numbers on the color bar represent different ranks of the density matrixes of the random states. Both axes are dimensionless.}
    \label{f3}
  \end{figure}

  In Fig. \ref{f3}, we plot how the maximum steering inequality violation changes in regard to negativity to the sixth power for ${10^5}$ Haar randomly generated three-qubit mixed states when ${\cal S}(\rho ) \geqslant 0$.
  \textcolor[rgb]{0.00,0.00,0.00}{We can see that state $|\psi \rangle _m$ is located at the upper boundary (blue circles)}, suggesting that its negativity and maximum steering inequality violation satisfy the relation in Eq. (\ref{Eq.53a}). The random states are under the boundary line, which means that their negativity and maximum steering inequality violation satisfy the inequality in Eq. (\ref{Eq.57a}). The results show that a complementary relation between negativity and the maximum steering inequality violation exists for arbitrary three-qubit states. Also, the higher the rank of the density matrix of the random state, the closer it is to the origin.

\begin{figure}[htbp]
\centering
\subfigure{\includegraphics[height=6.4cm]{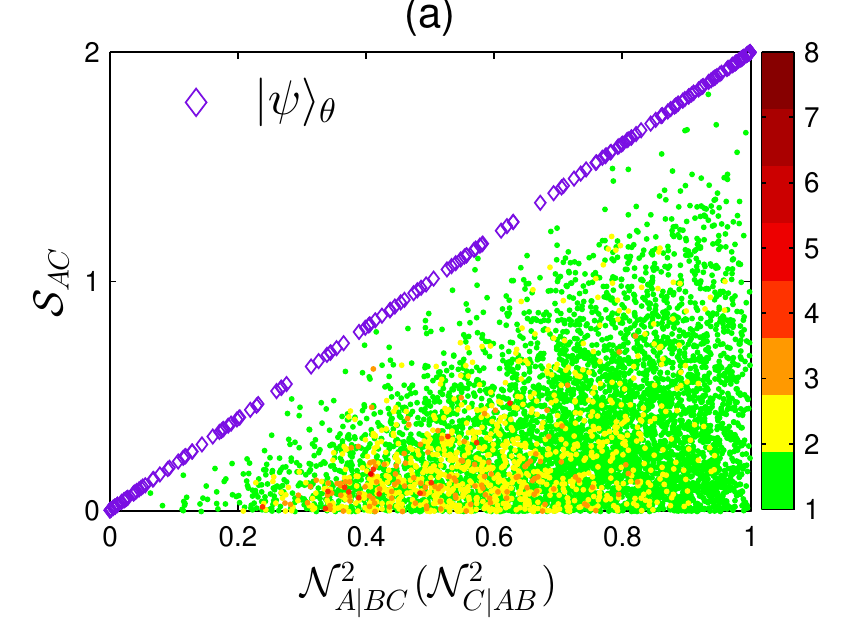}}
\subfigure{\includegraphics[height=6.4cm]{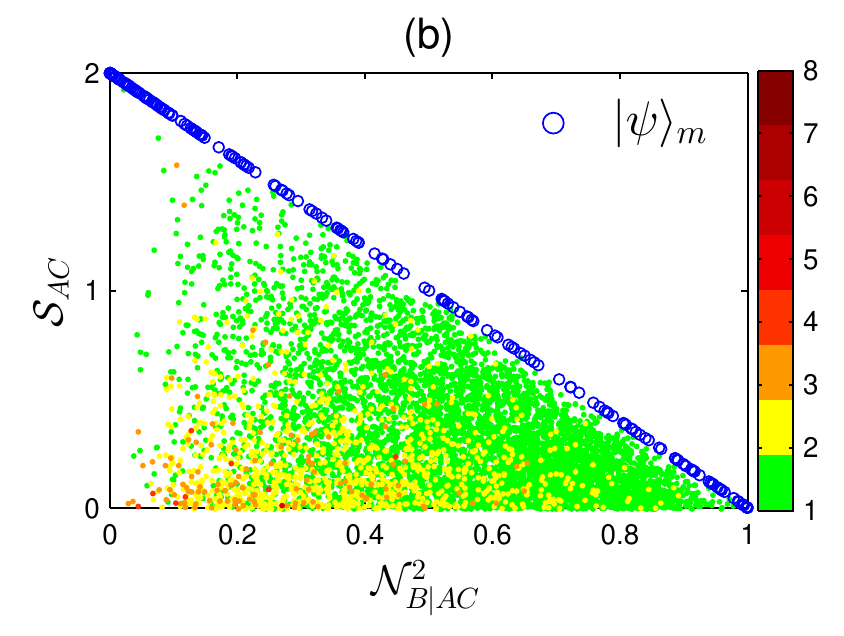}}
\caption{\textcolor[rgb]{0.00,0.00,0.00}{(color online). Relations between the pairwise steering inequality violation ${\cal S}_{AC}$ and three bipartite entanglement measures: (a) ${\cal N}_{A|BC}$ or ${\cal N}_{C|AB}$, (b) ${\cal N}_{B|AC}$. The numbers on the color bar represent different ranks of the random states. In (a), the boundary state is state $|\psi \rangle _{\theta}$ in Ref. \cite{PhysRevA.106.042415}. Both axes are dimensionless.}}
\label{f4}
\end{figure}

  \textcolor[rgb]{0.00,0.00,0.00}{
  It is worth mentioning that this complementary relation is obtained under the conditions of tripartite entanglement and the maximum pairwise steering inequality violation.
  Alternatively, is there an exact relation between pairwise steering inequality violation and bipartite entanglement measure in arbitrary three-qubit states?
  In the following, we take ${\cal S}_{AC}$ as an example, and investigate its relations with the bipartite entanglement measures ${\cal N}_{A|BC}$, ${\cal N}_{C|AB}$, and ${\cal N}_{B|AC}$.
   In Fig. \ref{f4}, we plot how ${\cal S}_{AC}$ changes with respect to ${\cal N}_{A|BC}$, ${\cal N}_{C|AB}$ and ${\cal N}_{B|AC}$, respectively.
   We find that the maximum of ${\cal S}_{AC}$ increases as ${\cal N}_{A|BC}$ (${\cal N}_{C|AB}$) increases. However, there exsits a complementary relation between ${\cal S}_{AC}$ and ${\cal N}_{B|AC}$.
   }

 \textcolor[rgb]{0.00,0.00,0.00}{
  {\it Corollary 1.\,}
  If an arbitrary three-qubit state $\rho_{ABC}$ has
  the same value of bipartite negativity ${\cal N}_{I|JK}(\rho)$ (${\rm{I,J,K}} \in \{ A,B,C\} $, ${\rm{I}} \ne {\rm{J}} \ne {\rm{K}}$) with state $|\psi \rangle _m$ (may need qubit permutations), the pairwise steering inequality violations
  of these two states satisfy the ordering ${\cal S}_{JK}(\rho) \leqslant {\cal S}_{JK}{(|\psi \rangle _m})$.
  The complementary relation of ${\cal N}_{I|JK}(\rho)$ and ${\cal S}_{JK}(\rho)$ is given by
  \begin{align}
    2{\cal N}_{I|JK}^2(\rho) + {\cal S}_{JK}(\rho) \leqslant 2.
    \label{Eq.71o}
  \end{align}
  }

  \textcolor[rgb]{0.00,0.00,0.00}{
  The proof is similar to the proof of the Theorem 3. The interpretation is that the increase of bipartite entanglement ${\cal N}_{I|JK}(\rho)$ decrease pairwise steering ${\cal S}_{JK}(\rho)$ by diminishing the entanglement of subsystem $\rho_{JK}$. Note that state $|\psi \rangle _m$ corresponds to maximum pairwise steering inequality violation ${\cal S}_{AC}$ for a fixed amount of bipartite negativity ${\cal N}_{B|AC}$. For the complementary relation between ${\cal S}_{AB}$ (${\cal S}_{BC}$) and ${\cal N}_{C|AB}$ (${\cal N}_{A|BC}$), the boundary state is the state after permutation of the latter (first) two qubits of state $|\psi \rangle _m$.
  }

  \textcolor[rgb]{0.00,0.00,0.00}{
    \section{negativity  versus maximum Bell-inequality violation\label{sec6}}
     {\it Theorem 4.\,}
    If an arbitrary three-qubit state $\rho$ has
    the same value of negativity with state $|\psi \rangle _m$ ($i,j,k $), the maximum Bell-inequality violation
    of these two states satisfy the ordering $  {\cal B}(\rho ) \leqslant {\cal B}{(|\psi \rangle _m})$.
    The complementary relation of negativity and the maximum Bell-inequality violation is given by
    \begin{align}
      {\cal N}{\left( {\rho } \right)^6} + {\cal B}\left( {\rho } \right) \leqslant 1.
      \label{Eq.70h}
    \end{align}
  }
  \begin{proof}
    \textcolor[rgb]{0.00,0.00,0.00}{
    To begin with, we assume that ${\cal B}(|\psi \rangle)  = {{\cal B}_{AB}}$, where $|\psi \rangle$ is a three-qubit pure state. For bipartite subsystem $\rho_{AB}$ of $|\psi \rangle$, there is a complementary relation between first-order coherence and maximum Bell-CHSH value \cite{PhysRevLett.115.220501}
    \begin{align}
      \frac{{\cal D}_{A B}^{2}}{2}+\left(\frac{{\cal B}_{AB}^{'}}{2 \sqrt{2}}\right)^{2} \leqslant \operatorname{Tr}\left(\rho_{A B}^{2}\right)-2\left(\varepsilon_{1} \varepsilon_{4}+\varepsilon_{2} \varepsilon_{3}\right),
      \label{Eq.71h}
    \end{align}
    where
    \begin{align}
      {\cal D}_{A B} =\sqrt{ \frac{\left(\mathcal{D} _{A}^{2}+\mathcal{D} _{B}^{2}\right)}{2}  }.
      \label{Eq.72h}
    \end{align}
    ${\cal D}_{A B} $ is the bipartite first-order coherence, and $\varepsilon_{1} \geqslant \varepsilon_{2} \geqslant \varepsilon_{3} \geqslant \varepsilon_{4}$ are the
    eigenvalues of $\rho_{AB}$. Here $\varepsilon_{3} = \varepsilon_{4}=0$ since $\rho_{AB}$ has the same eigenvalues as $\rho_C$, another subsystem of $|\psi \rangle$. If ${\cal B}_{AB}= 0$, Eq. (\ref{Eq.70h}) obviously holds. If ${\cal B}_{AB}>0$, from Eq. (\ref{Eq.17h}), we have
    \begin{align}
      {\cal B}_{AB}= M_{AB}-1.
      \label{Eq.73h}
    \end{align}
    Using Eqs. (\ref{Eq.16h}) and (\ref{Eq.73h}), Eq. (\ref{Eq.71h}) can be rewritten as
    \begin{align}
      {\cal D}_{A B}^{2}+{\cal B}_{AB}+1 \leqslant 2\operatorname{Tr}\left(\rho_{C}^{2}\right).
      \label{Eq.74h}
    \end{align}
    Then, from Eqs. (\ref{Eq.D1}) , (\ref{Eq.72h}) and (\ref{Eq.74h}), we get
    \begin{align}
      {\cal B}_{AB}\leqslant 2\operatorname{Tr}\left(\rho_{C}^{2}\right)- \operatorname{Tr}\left(\rho_{A}^{2}\right)-\operatorname{Tr}\left(\rho_{B}^{2}\right).
      \label{Eq.75h}
    \end{align}
    }

    \textcolor[rgb]{0.00,0.00,0.00}{
    On the other hand, from Ref. \cite{PhysRevLett.118.010401}, we can know that for any pure
    three-qubit state, the triple $(M_{AB},M_{AC},M_{BC})$ has the same ordering as $\left(s_{\text {iso }}^{A B}, s_{\text {iso }}^{A C}, s_{\text {iso }}^{B C}\right)$ of pairwise isotropic strengths, which happen to be a third of corresponding pairwise steering inequality violations. That means that the triple $({\cal B}_{AB},{\cal B}_{AC},{\cal B}_{BC})$ has the same ordering as $({\cal S}_{AB},{\cal S}_{AC},{\cal S}_{BC})$. From Eq. (\ref{Eq.B7}), we can obtain
    \begin{align}
      2\operatorname{Tr}\left(\rho_{C}^{2}\right)- \operatorname{Tr}\left(\rho_{A}^{2}\right)-\operatorname{Tr}\left(\rho_{B}^{2}\right)\leqslant 1- {2^6}\prod\limits_{\mu  = 1}^6 {{\lambda _\mu }}.
      \label{Eq.76h}
    \end{align}
    Therefore, it gives
    \begin{align}
      {\cal B}_{AB}\leqslant 1- {\left[ {2{{\left( {\prod\limits_i {\det {\rho _i}} } \right)}^{1/6}}} \right]^6}.
      \label{Eq.77h}
    \end{align}
    Substituting Eq. (\ref{Eq.5}) into Eq. (\ref{Eq.77h}), we obtain
    \begin{align}
      {\cal N}{\left( {\left| \psi  \right\rangle } \right)^6}+{\cal B}(|\psi \rangle)\leqslant 1.
      \label{Eq.78h}
    \end{align}
    Similarly, the above complementary relation also holds if
    ${{\cal B}_{AC}}$ or ${{\cal B}_{BC}}$ is the largest one among ${{\cal B}_{AB}}$ , ${{\cal B}_{AC}}$ and ${{\cal B}_{BC}}$.
    }

    \textcolor[rgb]{0.00,0.00,0.00}{
      The maximum Bell-inequality violation of state $|\psi \rangle _{m}$, from Eq. (\ref{Eq.16a}), is given by
      \begin{align}
        {\cal B}{(|\psi \rangle _m}) = \frac{{ 4{m^2}}}{{{{(1 + {m^2})}^2}}}.
        \label{Eq.79h}
      \end{align}
      Using Eqs. (\ref{Eq.40}) and (\ref{Eq.79h}), we have
      \begin{align}
        {\cal N}{\left( {\left| \psi  \right\rangle }_m \right)^6}+{\cal B}(|\psi \rangle_m)= 1,
        \label{Eq.80h}
      \end{align}
      which imply that the state $|\psi \rangle _{m}$ is the upper boundary states.
    }

    \textcolor[rgb]{0.00,0.00,0.00}{
      Furthermore, for an  arbitrary  three-qubit  state $\rho$, the maximum Bell-inequality violation is also a convex function \cite{PhysRevA.94.052126}, then we can extend the complementary relation between negativity and the maximum Bell-inequality violation to mixed states by a similar derivation to Eq. (\ref{Eq.44}). Thus, we have
      \begin{align}
        {\cal N}{\left( {\rho } \right)^6} + {\cal B}\left( {\rho } \right) \leqslant 1.
        \label{Eq.81h}
      \end{align}
    }
  \end{proof}

  \begin{figure}
      \centering
      \includegraphics[width=8.4cm]{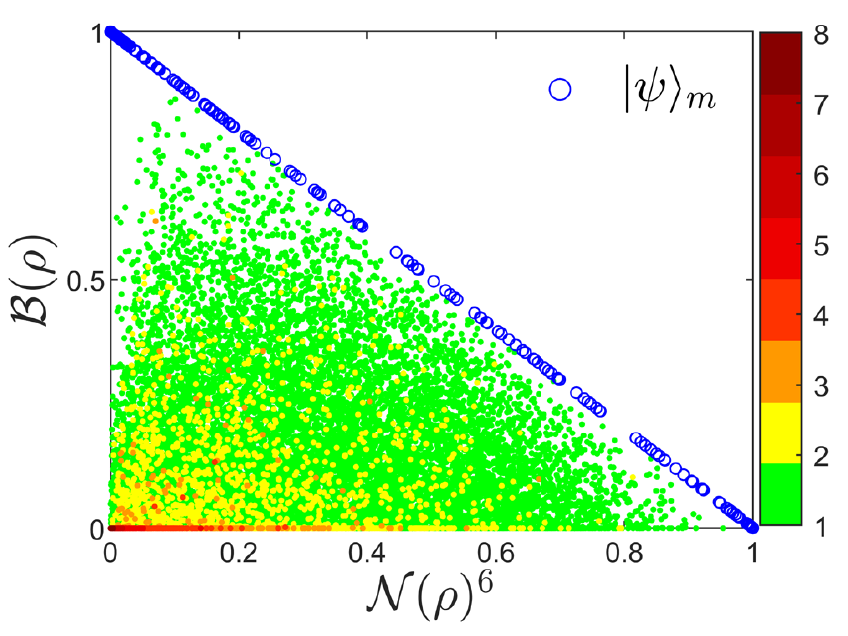}
      \caption{(Color online)  \textcolor[rgb]{0.00,0.00,0.00}{Complementary relation between negativity ${\cal N}(\rho )$ and the maximum Bell-inequality violation ${\cal B}(\rho )$ for ${10^5}$ Haar randomly generated three-qubit mixed states. The blue circles lie at the upper boundary with state $|\psi \rangle _m$. The numbers on the color bar represent different ranks of the random states. Both axes are dimensionless.}}
      \label{f5}
    \end{figure}

  \textcolor[rgb]{0.00,0.00,0.00}{
    In Fig. \ref{f5}, we plot the relation between negativity and the maximum Bell-inequality violation for ${10^5}$ Haar randomly generated three-qubit mixed states. We can see that
    state $|\psi \rangle _m$ is located at the upper boundary (blue circles), suggesting that its negativity and maximum Bell-inequality violation satisfy the complementary relation in Eq. (\ref{Eq.81h}). In particular, we find that the three-qubit state is hard to violate the Bell inequality when its rank is greater than three.
  }

\textcolor[rgb]{0.00,0.00,0.00}{
Also, if considering three pairwise Bell-inequality violations separately, we can obtain the complementary relations between pairwise Bell-inequality violations and bipartite entanglement in tripartite system.
}

\textcolor[rgb]{0.00,0.00,0.00}{
{\it Corollary 2.\,}
If an arbitrary three-qubit state $\rho_{ABC}$ has
  the same value of bipartite negativity ${\cal N}_{I|JK}(\rho)$ (${\rm{I,J,K}} \in \{ A,B,C\} $, ${\rm{I}} \ne {\rm{J}} \ne {\rm{K}}$) with state $|\psi \rangle _m$ (may need qubit permutations), the pairwise Bell-inequality violations
  of these two states satisfy the ordering ${\cal B}_{JK}(\rho) \leqslant {\cal B}_{JK}{(|\psi \rangle _m})$.
  The complementary relation of ${\cal N}_{I|JK}(\rho)$ and ${\cal B}_{JK}(\rho)$ is obtained as
  \begin{align}
    {\cal N}_{I|JK}^2(\rho) + {\cal B}_{JK}(\rho) \leqslant 1.
    \label{Eq.84o}
  \end{align}
}

  \section{conclusion\label{sec7}}

  In this paper, we found that exact complementary relations among entanglement, coherence, steering inequality violation \textcolor[rgb]{0.00,0.00,0.00}{and Bell nonlocality} exist for arbitrary three-qubit states.
  First of all, it was shown that the negativity is exactly the same as the GBC for the three-qubit pure states, although the negativity was always less than or equal to the GBC for three-qubit mixed states. Then the complementary relation between negativity and first-order coherence was established. For the three-qubit pure states, the first-order coherence is constrained to a range formed by two inequalities for a fixed amount of negativity. The upper boundary states of the complementary relation is state $|\psi \rangle _\alpha$, and $|\psi \rangle _m$ is the lower boundary state. For the three-qubit mixed states, the upper boundary is still valid while the lower boundary is ineffective. We can obtain that the higher the rank of the density matrix of the random state, the closer it is to the origin. Moreover, we investigated the complementary relation between the negativity and the maximum steering inequality violation for the three-qubit states. Interestingly, the $|\psi \rangle _m$ state takes the maximum steering inequality violation for a given negativity.
\textcolor[rgb]{0.00,0.00,0.00}{
  At last, we obtained that state $|\psi \rangle _m$ is also the
  upper boundary state of the complementary relation between negativity and the maximum Bell-inequality violation.}
These results show that these three quantum resources are closely related and can be transformed from one another.
\textcolor[rgb]{0.00,0.00,0.00}{
  In particular, our boundaries are useful for quantifying the maximum value of one resource can be converted from the other.
}
\begin{acknowledgements}
  This work was supported by the National Science Foundation of China (Grant Nos. 12004006, 12075001, and 12175001), Anhui Provincial Key Research and Development Plan (Grant No. 2022b13020004), and Anhui Provincial Natural Science Foundation (Grant No. 2008085QA43).
\end{acknowledgements}


\begin{thebibliography}{0}%
\makeatletter
\providecommand \@ifxundefined [1]{%
 \@ifx{#1\undefined}
}%
\providecommand \@ifnum [1]{%
 \ifnum #1\expandafter \@firstoftwo
 \else \expandafter \@secondoftwo
 \fi
}%
\providecommand \@ifx [1]{%
 \ifx #1\expandafter \@firstoftwo
 \else \expandafter \@secondoftwo
 \fi
}%
\providecommand \natexlab [1]{#1}%
\providecommand \enquote  [1]{``#1''}%
\providecommand \bibnamefont  [1]{#1}%
\providecommand \bibfnamefont [1]{#1}%
\providecommand \citenamefont [1]{#1}%
\providecommand \href@noop [0]{\@secondoftwo}%
\providecommand \href [0]{\begingroup \@sanitize@url \@href}%
\providecommand \@href[1]{\@@startlink{#1}\@@href}%
\providecommand \@@href[1]{\endgroup#1\@@endlink}%
\providecommand \@sanitize@url [0]{\catcode `\\12\catcode `\$12\catcode
  `\&12\catcode `\#12\catcode `\^12\catcode `\_12\catcode `\%12\relax}%
\providecommand \@@startlink[1]{}%
\providecommand \@@endlink[0]{}%
\providecommand \url  [0]{\begingroup\@sanitize@url \@url }%
\providecommand \@url [1]{\endgroup\@href {#1}{\urlprefix }}%
\providecommand \urlprefix  [0]{URL }%
\providecommand \Eprint [0]{\href }%
\providecommand \doibase [0]{http://dx.doi.org/}%
\providecommand \selectlanguage [0]{\@gobble}%
\providecommand \bibinfo  [0]{\@secondoftwo}%
\providecommand \bibfield  [0]{\@secondoftwo}%
\providecommand \translation [1]{[#1]}%
\providecommand \BibitemOpen [0]{}%
\providecommand \bibitemStop [0]{}%
\providecommand \bibitemNoStop [0]{.\EOS\space}%
\providecommand \EOS [0]{\spacefactor3000\relax}%
\providecommand \BibitemShut  [1]{\csname bibitem#1\endcsname}%
\let\auto@bib@innerbib\@empty
\end{thebibliography}%


\begin{thebibliography}{65}%
\makeatletter
\providecommand \@ifxundefined [1]{%
 \@ifx{#1\undefined}
}%
\providecommand \@ifnum [1]{%
 \ifnum #1\expandafter \@firstoftwo
 \else \expandafter \@secondoftwo
 \fi
}%
\providecommand \@ifx [1]{%
 \ifx #1\expandafter \@firstoftwo
 \else \expandafter \@secondoftwo
 \fi
}%
\providecommand \natexlab [1]{#1}%
\providecommand \enquote  [1]{``#1''}%
\providecommand \bibnamefont  [1]{#1}%
\providecommand \bibfnamefont [1]{#1}%
\providecommand \citenamefont [1]{#1}%
\providecommand \href@noop [0]{\@secondoftwo}%
\providecommand \href [0]{\begingroup \@sanitize@url \@href}%
\providecommand \@href[1]{\@@startlink{#1}\@@href}%
\providecommand \@@href[1]{\endgroup#1\@@endlink}%
\providecommand \@sanitize@url [0]{\catcode `\\12\catcode `\$12\catcode
  `\&12\catcode `\#12\catcode `\^12\catcode `\_12\catcode `\%12\relax}%
\providecommand \@@startlink[1]{}%
\providecommand \@@endlink[0]{}%
\providecommand \url  [0]{\begingroup\@sanitize@url \@url }%
\providecommand \@url [1]{\endgroup\@href {#1}{\urlprefix }}%
\providecommand \urlprefix  [0]{URL }%
\providecommand \Eprint [0]{\href }%
\providecommand \doibase [0]{http://dx.doi.org/}%
\providecommand \selectlanguage [0]{\@gobble}%
\providecommand \bibinfo  [0]{\@secondoftwo}%
\providecommand \bibfield  [0]{\@secondoftwo}%
\providecommand \translation [1]{[#1]}%
\providecommand \BibitemOpen [0]{}%
\providecommand \bibitemStop [0]{}%
\providecommand \bibitemNoStop [0]{.\EOS\space}%
\providecommand \EOS [0]{\spacefactor3000\relax}%
\providecommand \BibitemShut  [1]{\csname bibitem#1\endcsname}%
\let\auto@bib@innerbib\@empty
\bibitem [{\citenamefont {Ekert}(1991)}]{PhysRevLett.67.661}%
  \BibitemOpen
  \bibfield  {author} {\bibinfo {author} {\bibfnamefont {A.~K.}\ \bibnamefont
  {Ekert}},\ }\href {\doibase 10.1103/PhysRevLett.67.661} {\bibfield  {journal}
  {\bibinfo  {journal} {Phys. Rev. Lett.}\ }\textbf {\bibinfo {volume} {67}},\
  \bibinfo {pages} {661} (\bibinfo {year} {1991})}\BibitemShut {NoStop}%
\bibitem [{\citenamefont {Bennett}\ \emph {et~al.}(1993)\citenamefont
  {Bennett}, \citenamefont {Brassard}, \citenamefont {Cr\'epeau}, \citenamefont
  {Jozsa}, \citenamefont {Peres},\ and\ \citenamefont
  {Wootters}}]{PhysRevLett.70.1895}%
  \BibitemOpen
  \bibfield  {author} {\bibinfo {author} {\bibfnamefont {C.~H.}\ \bibnamefont
  {Bennett}}, \bibinfo {author} {\bibfnamefont {G.}~\bibnamefont {Brassard}},
  \bibinfo {author} {\bibfnamefont {C.}~\bibnamefont {Cr\'epeau}}, \bibinfo
  {author} {\bibfnamefont {R.}~\bibnamefont {Jozsa}}, \bibinfo {author}
  {\bibfnamefont {A.}~\bibnamefont {Peres}}, \ and\ \bibinfo {author}
  {\bibfnamefont {W.~K.}\ \bibnamefont {Wootters}},\ }\href {\doibase
  10.1103/PhysRevLett.70.1895} {\bibfield  {journal} {\bibinfo  {journal}
  {Phys. Rev. Lett.}\ }\textbf {\bibinfo {volume} {70}},\ \bibinfo {pages}
  {1895} (\bibinfo {year} {1993})}\BibitemShut {NoStop}%
\bibitem [{\citenamefont {Bennett}\ and\ \citenamefont
  {Wiesner}(1992)}]{PhysRevLett.69.2881}%
  \BibitemOpen
  \bibfield  {author} {\bibinfo {author} {\bibfnamefont {C.~H.}\ \bibnamefont
  {Bennett}}\ and\ \bibinfo {author} {\bibfnamefont {S.~J.}\ \bibnamefont
  {Wiesner}},\ }\href {\doibase 10.1103/PhysRevLett.69.2881} {\bibfield
  {journal} {\bibinfo  {journal} {Phys. Rev. Lett.}\ }\textbf {\bibinfo
  {volume} {69}},\ \bibinfo {pages} {2881} (\bibinfo {year}
  {1992})}\BibitemShut {NoStop}%
\bibitem [{\citenamefont {Wootters}(1998)}]{PhysRevLett.80.2245}%
  \BibitemOpen
  \bibfield  {author} {\bibinfo {author} {\bibfnamefont {W.~K.}\ \bibnamefont
  {Wootters}},\ }\href {\doibase 10.1103/PhysRevLett.80.2245} {\bibfield
  {journal} {\bibinfo  {journal} {Phys. Rev. Lett.}\ }\textbf {\bibinfo
  {volume} {80}},\ \bibinfo {pages} {2245} (\bibinfo {year}
  {1998})}\BibitemShut {NoStop}%
\bibitem [{\citenamefont {Eisert}\ and\ \citenamefont
  {Plenio}(1999)}]{doi:10.1080/09500349908231260}%
  \BibitemOpen
  \bibfield  {author} {\bibinfo {author} {\bibfnamefont {J.}~\bibnamefont
  {Eisert}}\ and\ \bibinfo {author} {\bibfnamefont {M.~B.}\ \bibnamefont
  {Plenio}},\ }\href {\doibase 10.1080/09500349908231260} {\bibfield  {journal}
  {\bibinfo  {journal} {J Mod Opt.}\ }\textbf {\bibinfo {volume} {46}},\
  \bibinfo {pages} {145} (\bibinfo {year} {1999})}\BibitemShut {NoStop}%
\bibitem [{\citenamefont {Vidal}\ and\ \citenamefont
  {Werner}(2002)}]{PhysRevA.65.032314}%
  \BibitemOpen
  \bibfield  {author} {\bibinfo {author} {\bibfnamefont {G.}~\bibnamefont
  {Vidal}}\ and\ \bibinfo {author} {\bibfnamefont {R.~F.}\ \bibnamefont
  {Werner}},\ }\href {\doibase 10.1103/PhysRevA.65.032314} {\bibfield
  {journal} {\bibinfo  {journal} {Phys. Rev. A}\ }\textbf {\bibinfo {volume}
  {65}},\ \bibinfo {pages} {032314} (\bibinfo {year} {2002})}\BibitemShut
  {NoStop}%
\bibitem [{\citenamefont {Sab{\'\i}n}\ and\ \citenamefont
  {Garc{\'\i}a-Alcaine}(2008)}]{sabin2008classification}%
  \BibitemOpen
  \bibfield  {author} {\bibinfo {author} {\bibfnamefont {C.}~\bibnamefont
  {Sab{\'\i}n}}\ and\ \bibinfo {author} {\bibfnamefont {G.}~\bibnamefont
  {Garc{\'\i}a-Alcaine}},\ }\href
  {https://link.springer.com/article/10.1140/epjd/e2008-00112-5} {\bibfield
  {journal} {\bibinfo  {journal} {Eur. Phys. J. D}\ }\textbf {\bibinfo {volume}
  {48}},\ \bibinfo {pages} {435} (\bibinfo {year} {2008})}\BibitemShut
  {NoStop}%
\bibitem [{\citenamefont {Li}\ and\ \citenamefont
  {Shang}(2022)}]{PhysRevResearch.4.023059}%
  \BibitemOpen
  \bibfield  {author} {\bibinfo {author} {\bibfnamefont {Y.}~\bibnamefont
  {Li}}\ and\ \bibinfo {author} {\bibfnamefont {J.}~\bibnamefont {Shang}},\
  }\href {\doibase 10.1103/PhysRevResearch.4.023059} {\bibfield  {journal}
  {\bibinfo  {journal} {Phys. Rev. Research}\ }\textbf {\bibinfo {volume}
  {4}},\ \bibinfo {pages} {023059} (\bibinfo {year} {2022})}\BibitemShut
  {NoStop}%
\bibitem [{\citenamefont {Gour}(2005)}]{PhysRevA.71.012318}%
  \BibitemOpen
  \bibfield  {author} {\bibinfo {author} {\bibfnamefont {G.}~\bibnamefont
  {Gour}},\ }\href {\doibase 10.1103/PhysRevA.71.012318} {\bibfield  {journal}
  {\bibinfo  {journal} {Phys. Rev. A}\ }\textbf {\bibinfo {volume} {71}},\
  \bibinfo {pages} {012318} (\bibinfo {year} {2005})}\BibitemShut {NoStop}%
\bibitem [{\citenamefont {Miranowicz}\ and\ \citenamefont
  {Grudka}(2004{\natexlab{a}})}]{PhysRevA.70.032326}%
  \BibitemOpen
  \bibfield  {author} {\bibinfo {author} {\bibfnamefont {A.}~\bibnamefont
  {Miranowicz}}\ and\ \bibinfo {author} {\bibfnamefont {A.}~\bibnamefont
  {Grudka}},\ }\href {\doibase 10.1103/PhysRevA.70.032326} {\bibfield
  {journal} {\bibinfo  {journal} {Phys. Rev. A}\ }\textbf {\bibinfo {volume}
  {70}},\ \bibinfo {pages} {032326} (\bibinfo {year}
  {2004}{\natexlab{a}})}\BibitemShut {NoStop}%
\bibitem [{\citenamefont {Miranowicz}\ and\ \citenamefont
  {Grudka}(2004{\natexlab{b}})}]{miranowicz2004comparative}%
  \BibitemOpen
  \bibfield  {author} {\bibinfo {author} {\bibfnamefont {A.}~\bibnamefont
  {Miranowicz}}\ and\ \bibinfo {author} {\bibfnamefont {A.}~\bibnamefont
  {Grudka}},\ }\href
  {https://iopscience.iop.org/article/10.1088/1464-4266/6/12/009} {\bibfield
  {journal} {\bibinfo  {journal} {J. Opt.}\ }\textbf {\bibinfo {volume} {6}},\
  \bibinfo {pages} {542} (\bibinfo {year} {2004}{\natexlab{b}})}\BibitemShut
  {NoStop}%
\bibitem [{\citenamefont {Eltschka}\ \emph {et~al.}(2015)\citenamefont
  {Eltschka}, \citenamefont {T\'oth},\ and\ \citenamefont
  {Siewert}}]{PhysRevA.91.032327}%
  \BibitemOpen
  \bibfield  {author} {\bibinfo {author} {\bibfnamefont {C.}~\bibnamefont
  {Eltschka}}, \bibinfo {author} {\bibfnamefont {G.}~\bibnamefont {T\'oth}}, \
  and\ \bibinfo {author} {\bibfnamefont {J.}~\bibnamefont {Siewert}},\ }\href
  {\doibase 10.1103/PhysRevA.91.032327} {\bibfield  {journal} {\bibinfo
  {journal} {Phys. Rev. A}\ }\textbf {\bibinfo {volume} {91}},\ \bibinfo
  {pages} {032327} (\bibinfo {year} {2015})}\BibitemShut {NoStop}%
\bibitem [{\citenamefont {Dong}\ \emph {et~al.}(2022)\citenamefont {Dong},
  \citenamefont {Wei}, \citenamefont {Song}, \citenamefont {Wang},\ and\
  \citenamefont {Ye}}]{PhysRevA.106.042415}%
  \BibitemOpen
  \bibfield  {author} {\bibinfo {author} {\bibfnamefont {D.-D.}\ \bibnamefont
  {Dong}}, \bibinfo {author} {\bibfnamefont {G.-B.}\ \bibnamefont {Wei}},
  \bibinfo {author} {\bibfnamefont {X.-K.}\ \bibnamefont {Song}}, \bibinfo
  {author} {\bibfnamefont {D.}~\bibnamefont {Wang}}, \ and\ \bibinfo {author}
  {\bibfnamefont {L.}~\bibnamefont {Ye}},\ }\href {\doibase
  10.1103/PhysRevA.106.042415} {\bibfield  {journal} {\bibinfo  {journal}
  {Phys. Rev. A}\ }\textbf {\bibinfo {volume} {106}},\ \bibinfo {pages}
  {042415} (\bibinfo {year} {2022})}\BibitemShut {NoStop}%
\bibitem [{\citenamefont {Nielsen}\ and\ \citenamefont
  {Chuang}(2010)}]{nielsen_chuang_2010}%
  \BibitemOpen
  \bibfield  {author} {\bibinfo {author} {\bibfnamefont {M.~A.}\ \bibnamefont
  {Nielsen}}\ and\ \bibinfo {author} {\bibfnamefont {I.~L.}\ \bibnamefont
  {Chuang}},\ }\href@noop {} {\emph {\bibinfo {title} {Quantum Computation and
  Quantum Information: 10th Anniversary Edition}}}\ (\bibinfo  {publisher}
  {Cambridge University Press},\ \bibinfo {year} {2010})\BibitemShut {NoStop}%
\bibitem [{\citenamefont {Streltsov}\ \emph {et~al.}(2017)\citenamefont
  {Streltsov}, \citenamefont {Adesso},\ and\ \citenamefont
  {Plenio}}]{RevModPhys.89.041003}%
  \BibitemOpen
  \bibfield  {author} {\bibinfo {author} {\bibfnamefont {A.}~\bibnamefont
  {Streltsov}}, \bibinfo {author} {\bibfnamefont {G.}~\bibnamefont {Adesso}}, \
  and\ \bibinfo {author} {\bibfnamefont {M.~B.}\ \bibnamefont {Plenio}},\
  }\href {\doibase 10.1103/RevModPhys.89.041003} {\bibfield  {journal}
  {\bibinfo  {journal} {Rev. Mod. Phys.}\ }\textbf {\bibinfo {volume} {89}},\
  \bibinfo {pages} {041003} (\bibinfo {year} {2017})}\BibitemShut {NoStop}%
\bibitem [{\citenamefont {Baumgratz}\ \emph {et~al.}(2014)\citenamefont
  {Baumgratz}, \citenamefont {Cramer},\ and\ \citenamefont
  {Plenio}}]{PhysRevLett.113.140401}%
  \BibitemOpen
  \bibfield  {author} {\bibinfo {author} {\bibfnamefont {T.}~\bibnamefont
  {Baumgratz}}, \bibinfo {author} {\bibfnamefont {M.}~\bibnamefont {Cramer}}, \
  and\ \bibinfo {author} {\bibfnamefont {M.~B.}\ \bibnamefont {Plenio}},\
  }\href {\doibase 10.1103/PhysRevLett.113.140401} {\bibfield  {journal}
  {\bibinfo  {journal} {Phys. Rev. Lett.}\ }\textbf {\bibinfo {volume} {113}},\
  \bibinfo {pages} {140401} (\bibinfo {year} {2014})}\BibitemShut {NoStop}%
\bibitem [{\citenamefont {Winter}\ and\ \citenamefont
  {Yang}(2016)}]{PhysRevLett.116.120404}%
  \BibitemOpen
  \bibfield  {author} {\bibinfo {author} {\bibfnamefont {A.}~\bibnamefont
  {Winter}}\ and\ \bibinfo {author} {\bibfnamefont {D.}~\bibnamefont {Yang}},\
  }\href {\doibase 10.1103/PhysRevLett.116.120404} {\bibfield  {journal}
  {\bibinfo  {journal} {Phys. Rev. Lett.}\ }\textbf {\bibinfo {volume} {116}},\
  \bibinfo {pages} {120404} (\bibinfo {year} {2016})}\BibitemShut {NoStop}%
\bibitem [{\citenamefont {Girolami}(2014)}]{PhysRevLett.113.170401}%
  \BibitemOpen
  \bibfield  {author} {\bibinfo {author} {\bibfnamefont {D.}~\bibnamefont
  {Girolami}},\ }\href {\doibase 10.1103/PhysRevLett.113.170401} {\bibfield
  {journal} {\bibinfo  {journal} {Phys. Rev. Lett.}\ }\textbf {\bibinfo
  {volume} {113}},\ \bibinfo {pages} {170401} (\bibinfo {year}
  {2014})}\BibitemShut {NoStop}%
\bibitem [{\citenamefont {Brand\~ao}\ \emph {et~al.}(2013)\citenamefont
  {Brand\~ao}, \citenamefont {Horodecki}, \citenamefont {Oppenheim},
  \citenamefont {Renes},\ and\ \citenamefont
  {Spekkens}}]{PhysRevLett.111.250404}%
  \BibitemOpen
  \bibfield  {author} {\bibinfo {author} {\bibfnamefont {F.~G. S.~L.}\
  \bibnamefont {Brand\~ao}}, \bibinfo {author} {\bibfnamefont {M.}~\bibnamefont
  {Horodecki}}, \bibinfo {author} {\bibfnamefont {J.}~\bibnamefont
  {Oppenheim}}, \bibinfo {author} {\bibfnamefont {J.~M.}\ \bibnamefont
  {Renes}}, \ and\ \bibinfo {author} {\bibfnamefont {R.~W.}\ \bibnamefont
  {Spekkens}},\ }\href {\doibase 10.1103/PhysRevLett.111.250404} {\bibfield
  {journal} {\bibinfo  {journal} {Phys. Rev. Lett.}\ }\textbf {\bibinfo
  {volume} {111}},\ \bibinfo {pages} {250404} (\bibinfo {year}
  {2013})}\BibitemShut {NoStop}%
\bibitem [{\citenamefont {Hu}\ and\ \citenamefont
  {Fan}(2016)}]{hu2016extracting}%
  \BibitemOpen
  \bibfield  {author} {\bibinfo {author} {\bibfnamefont {X.}~\bibnamefont
  {Hu}}\ and\ \bibinfo {author} {\bibfnamefont {H.}~\bibnamefont {Fan}},\
  }\href {https://www.nature.com/articles/srep34380} {\bibfield  {journal}
  {\bibinfo  {journal} {Sci. Rep.}\ }\textbf {\bibinfo {volume} {6}},\ \bibinfo
  {pages} {1} (\bibinfo {year} {2016})}\BibitemShut {NoStop}%
\bibitem [{\citenamefont {Gallego}\ and\ \citenamefont
  {Aolita}(2015)}]{PhysRevX.5.041008}%
  \BibitemOpen
  \bibfield  {author} {\bibinfo {author} {\bibfnamefont {R.}~\bibnamefont
  {Gallego}}\ and\ \bibinfo {author} {\bibfnamefont {L.}~\bibnamefont
  {Aolita}},\ }\href {\doibase 10.1103/PhysRevX.5.041008} {\bibfield  {journal}
  {\bibinfo  {journal} {Phys. Rev. X}\ }\textbf {\bibinfo {volume} {5}},\
  \bibinfo {pages} {041008} (\bibinfo {year} {2015})}\BibitemShut {NoStop}%
\bibitem [{\citenamefont {Cavalcanti}\ \emph {et~al.}(2009)\citenamefont
  {Cavalcanti}, \citenamefont {Jones}, \citenamefont {Wiseman},\ and\
  \citenamefont {Reid}}]{PhysRevA.80.032112}%
  \BibitemOpen
  \bibfield  {author} {\bibinfo {author} {\bibfnamefont {E.~G.}\ \bibnamefont
  {Cavalcanti}}, \bibinfo {author} {\bibfnamefont {S.~J.}\ \bibnamefont
  {Jones}}, \bibinfo {author} {\bibfnamefont {H.~M.}\ \bibnamefont {Wiseman}},
  \ and\ \bibinfo {author} {\bibfnamefont {M.~D.}\ \bibnamefont {Reid}},\
  }\href {\doibase 10.1103/PhysRevA.80.032112} {\bibfield  {journal} {\bibinfo
  {journal} {Phys. Rev. A}\ }\textbf {\bibinfo {volume} {80}},\ \bibinfo
  {pages} {032112} (\bibinfo {year} {2009})}\BibitemShut {NoStop}%
\bibitem [{\citenamefont {Costa}\ and\ \citenamefont
  {Angelo}(2016)}]{PhysRevA.93.020103}%
  \BibitemOpen
  \bibfield  {author} {\bibinfo {author} {\bibfnamefont {A.~C.~S.}\
  \bibnamefont {Costa}}\ and\ \bibinfo {author} {\bibfnamefont {R.~M.}\
  \bibnamefont {Angelo}},\ }\href {\doibase 10.1103/PhysRevA.93.020103}
  {\bibfield  {journal} {\bibinfo  {journal} {Phys. Rev. A}\ }\textbf {\bibinfo
  {volume} {93}},\ \bibinfo {pages} {020103} (\bibinfo {year}
  {2016})}\BibitemShut {NoStop}%
\bibitem [{\citenamefont {\ifmmode~\dot{Z}\else \.{Z}\fi{}ukowski}\ \emph
  {et~al.}(2015)\citenamefont {\ifmmode~\dot{Z}\else \.{Z}\fi{}ukowski},
  \citenamefont {Dutta},\ and\ \citenamefont {Yin}}]{PhysRevA.91.032107}%
  \BibitemOpen
  \bibfield  {author} {\bibinfo {author} {\bibfnamefont {M.}~\bibnamefont
  {\ifmmode~\dot{Z}\else \.{Z}\fi{}ukowski}}, \bibinfo {author} {\bibfnamefont
  {A.}~\bibnamefont {Dutta}}, \ and\ \bibinfo {author} {\bibfnamefont
  {Z.}~\bibnamefont {Yin}},\ }\href {\doibase 10.1103/PhysRevA.91.032107}
  {\bibfield  {journal} {\bibinfo  {journal} {Phys. Rev. A}\ }\textbf {\bibinfo
  {volume} {91}},\ \bibinfo {pages} {032107} (\bibinfo {year}
  {2015})}\BibitemShut {NoStop}%
\bibitem [{\citenamefont {Walborn}\ \emph {et~al.}(2011)\citenamefont
  {Walborn}, \citenamefont {Salles}, \citenamefont {Gomes}, \citenamefont
  {Toscano},\ and\ \citenamefont {Souto~Ribeiro}}]{PhysRevLett.106.130402}%
  \BibitemOpen
  \bibfield  {author} {\bibinfo {author} {\bibfnamefont {S.~P.}\ \bibnamefont
  {Walborn}}, \bibinfo {author} {\bibfnamefont {A.}~\bibnamefont {Salles}},
  \bibinfo {author} {\bibfnamefont {R.~M.}\ \bibnamefont {Gomes}}, \bibinfo
  {author} {\bibfnamefont {F.}~\bibnamefont {Toscano}}, \ and\ \bibinfo
  {author} {\bibfnamefont {P.~H.}\ \bibnamefont {Souto~Ribeiro}},\ }\href
  {\doibase 10.1103/PhysRevLett.106.130402} {\bibfield  {journal} {\bibinfo
  {journal} {Phys. Rev. Lett.}\ }\textbf {\bibinfo {volume} {106}},\ \bibinfo
  {pages} {130402} (\bibinfo {year} {2011})}\BibitemShut {NoStop}%
\bibitem [{\citenamefont {Schneeloch}\ \emph {et~al.}(2013)\citenamefont
  {Schneeloch}, \citenamefont {Broadbent}, \citenamefont {Walborn},
  \citenamefont {Cavalcanti},\ and\ \citenamefont
  {Howell}}]{PhysRevA.87.062103}%
  \BibitemOpen
  \bibfield  {author} {\bibinfo {author} {\bibfnamefont {J.}~\bibnamefont
  {Schneeloch}}, \bibinfo {author} {\bibfnamefont {C.~J.}\ \bibnamefont
  {Broadbent}}, \bibinfo {author} {\bibfnamefont {S.~P.}\ \bibnamefont
  {Walborn}}, \bibinfo {author} {\bibfnamefont {E.~G.}\ \bibnamefont
  {Cavalcanti}}, \ and\ \bibinfo {author} {\bibfnamefont {J.~C.}\ \bibnamefont
  {Howell}},\ }\href {\doibase 10.1103/PhysRevA.87.062103} {\bibfield
  {journal} {\bibinfo  {journal} {Phys. Rev. A}\ }\textbf {\bibinfo {volume}
  {87}},\ \bibinfo {pages} {062103} (\bibinfo {year} {2013})}\BibitemShut
  {NoStop}%
\bibitem [{\citenamefont {Costa}\ \emph {et~al.}(2018)\citenamefont {Costa},
  \citenamefont {Uola},\ and\ \citenamefont {G\"uhne}}]{PhysRevA.98.050104}%
  \BibitemOpen
  \bibfield  {author} {\bibinfo {author} {\bibfnamefont {A.~C.~S.}\
  \bibnamefont {Costa}}, \bibinfo {author} {\bibfnamefont {R.}~\bibnamefont
  {Uola}}, \ and\ \bibinfo {author} {\bibfnamefont {O.}~\bibnamefont
  {G\"uhne}},\ }\href {\doibase 10.1103/PhysRevA.98.050104} {\bibfield
  {journal} {\bibinfo  {journal} {Phys. Rev. A}\ }\textbf {\bibinfo {volume}
  {98}},\ \bibinfo {pages} {050104} (\bibinfo {year} {2018})}\BibitemShut
  {NoStop}%
\bibitem [{\citenamefont {Kriv\'achy}\ \emph {et~al.}(2018)\citenamefont
  {Kriv\'achy}, \citenamefont {Fr\"owis},\ and\ \citenamefont
  {Brunner}}]{PhysRevA.98.062111}%
  \BibitemOpen
  \bibfield  {author} {\bibinfo {author} {\bibfnamefont {T.}~\bibnamefont
  {Kriv\'achy}}, \bibinfo {author} {\bibfnamefont {F.}~\bibnamefont
  {Fr\"owis}}, \ and\ \bibinfo {author} {\bibfnamefont {N.}~\bibnamefont
  {Brunner}},\ }\href {\doibase 10.1103/PhysRevA.98.062111} {\bibfield
  {journal} {\bibinfo  {journal} {Phys. Rev. A}\ }\textbf {\bibinfo {volume}
  {98}},\ \bibinfo {pages} {062111} (\bibinfo {year} {2018})}\BibitemShut
  {NoStop}%
\bibitem [{\citenamefont {Branciard}\ and\ \citenamefont
  {Gisin}(2011)}]{PhysRevLett.107.020401}%
  \BibitemOpen
  \bibfield  {author} {\bibinfo {author} {\bibfnamefont {C.}~\bibnamefont
  {Branciard}}\ and\ \bibinfo {author} {\bibfnamefont {N.}~\bibnamefont
  {Gisin}},\ }\href {\doibase 10.1103/PhysRevLett.107.020401} {\bibfield
  {journal} {\bibinfo  {journal} {Phys. Rev. Lett.}\ }\textbf {\bibinfo
  {volume} {107}},\ \bibinfo {pages} {020401} (\bibinfo {year}
  {2011})}\BibitemShut {NoStop}%
\bibitem [{\citenamefont {Cavalcanti}\ \emph {et~al.}(2017)\citenamefont
  {Cavalcanti}, \citenamefont {Skrzypczyk},\ and\ \citenamefont {\ifmmode
  \check{S}\else \v{S}\fi{}upi\ifmmode~\acute{c}\else
  \'{c}\fi{}}}]{PhysRevLett.119.110501}%
  \BibitemOpen
  \bibfield  {author} {\bibinfo {author} {\bibfnamefont {D.}~\bibnamefont
  {Cavalcanti}}, \bibinfo {author} {\bibfnamefont {P.}~\bibnamefont
  {Skrzypczyk}}, \ and\ \bibinfo {author} {\bibfnamefont {I.}~\bibnamefont
  {\ifmmode \check{S}\else \v{S}\fi{}upi\ifmmode~\acute{c}\else \'{c}\fi{}}},\
  }\href {\doibase 10.1103/PhysRevLett.119.110501} {\bibfield  {journal}
  {\bibinfo  {journal} {Phys. Rev. Lett.}\ }\textbf {\bibinfo {volume} {119}},\
  \bibinfo {pages} {110501} (\bibinfo {year} {2017})}\BibitemShut {NoStop}%
\bibitem [{\citenamefont {Branciard}\ \emph {et~al.}(2012)\citenamefont
  {Branciard}, \citenamefont {Cavalcanti}, \citenamefont {Walborn},
  \citenamefont {Scarani},\ and\ \citenamefont {Wiseman}}]{PhysRevA.85.010301}%
  \BibitemOpen
  \bibfield  {author} {\bibinfo {author} {\bibfnamefont {C.}~\bibnamefont
  {Branciard}}, \bibinfo {author} {\bibfnamefont {E.~G.}\ \bibnamefont
  {Cavalcanti}}, \bibinfo {author} {\bibfnamefont {S.~P.}\ \bibnamefont
  {Walborn}}, \bibinfo {author} {\bibfnamefont {V.}~\bibnamefont {Scarani}}, \
  and\ \bibinfo {author} {\bibfnamefont {H.~M.}\ \bibnamefont {Wiseman}},\
  }\href {\doibase 10.1103/PhysRevA.85.010301} {\bibfield  {journal} {\bibinfo
  {journal} {Phys. Rev. A}\ }\textbf {\bibinfo {volume} {85}},\ \bibinfo
  {pages} {010301} (\bibinfo {year} {2012})}\BibitemShut {NoStop}%
\bibitem [{\citenamefont {Wang}\ \emph {et~al.}(2013)\citenamefont {Wang},
  \citenamefont {Bao}, \citenamefont {Li}, \citenamefont {Zhou},\ and\
  \citenamefont {Li}}]{PhysRevA.88.052322}%
  \BibitemOpen
  \bibfield  {author} {\bibinfo {author} {\bibfnamefont {Y.}~\bibnamefont
  {Wang}}, \bibinfo {author} {\bibfnamefont {W.-S.}\ \bibnamefont {Bao}},
  \bibinfo {author} {\bibfnamefont {H.-W.}\ \bibnamefont {Li}}, \bibinfo
  {author} {\bibfnamefont {C.}~\bibnamefont {Zhou}}, \ and\ \bibinfo {author}
  {\bibfnamefont {Y.}~\bibnamefont {Li}},\ }\href {\doibase
  10.1103/PhysRevA.88.052322} {\bibfield  {journal} {\bibinfo  {journal} {Phys.
  Rev. A}\ }\textbf {\bibinfo {volume} {88}},\ \bibinfo {pages} {052322}
  (\bibinfo {year} {2013})}\BibitemShut {NoStop}%
\bibitem [{\citenamefont {Kaur}\ \emph {et~al.}(2020)\citenamefont {Kaur},
  \citenamefont {Wilde},\ and\ \citenamefont {Winter}}]{Kaur_2020}%
  \BibitemOpen
  \bibfield  {author} {\bibinfo {author} {\bibfnamefont {E.}~\bibnamefont
  {Kaur}}, \bibinfo {author} {\bibfnamefont {M.~M.}\ \bibnamefont {Wilde}}, \
  and\ \bibinfo {author} {\bibfnamefont {A.}~\bibnamefont {Winter}},\ }\href
  {\doibase 10.1088/1367-2630/ab6eaa} {\bibfield  {journal} {\bibinfo
  {journal} {New J. Phys.}\ }\textbf {\bibinfo {volume} {22}},\ \bibinfo
  {pages} {023039} (\bibinfo {year} {2020})}\BibitemShut {NoStop}%
\bibitem [{\citenamefont {Piani}\ and\ \citenamefont
  {Watrous}(2015)}]{PhysRevLett.114.060404}%
  \BibitemOpen
  \bibfield  {author} {\bibinfo {author} {\bibfnamefont {M.}~\bibnamefont
  {Piani}}\ and\ \bibinfo {author} {\bibfnamefont {J.}~\bibnamefont
  {Watrous}},\ }\href {\doibase 10.1103/PhysRevLett.114.060404} {\bibfield
  {journal} {\bibinfo  {journal} {Phys. Rev. Lett.}\ }\textbf {\bibinfo
  {volume} {114}},\ \bibinfo {pages} {060404} (\bibinfo {year}
  {2015})}\BibitemShut {NoStop}%
\bibitem [{\citenamefont {Uola}\ \emph {et~al.}(2019)\citenamefont {Uola},
  \citenamefont {Kraft}, \citenamefont {Shang}, \citenamefont {Yu},\ and\
  \citenamefont {G\"uhne}}]{PhysRevLett.122.130404}%
  \BibitemOpen
  \bibfield  {author} {\bibinfo {author} {\bibfnamefont {R.}~\bibnamefont
  {Uola}}, \bibinfo {author} {\bibfnamefont {T.}~\bibnamefont {Kraft}},
  \bibinfo {author} {\bibfnamefont {J.}~\bibnamefont {Shang}}, \bibinfo
  {author} {\bibfnamefont {X.-D.}\ \bibnamefont {Yu}}, \ and\ \bibinfo {author}
  {\bibfnamefont {O.}~\bibnamefont {G\"uhne}},\ }\href {\doibase
  10.1103/PhysRevLett.122.130404} {\bibfield  {journal} {\bibinfo  {journal}
  {Phys. Rev. Lett.}\ }\textbf {\bibinfo {volume} {122}},\ \bibinfo {pages}
  {130404} (\bibinfo {year} {2019})}\BibitemShut {NoStop}%
\bibitem [{\citenamefont {Svozil\'{\i}k}\ \emph {et~al.}(2015)\citenamefont
  {Svozil\'{\i}k}, \citenamefont {Vall\'es}, \citenamefont
  {Pe\ifmmode~\check{r}\else \v{r}\fi{}ina},\ and\ \citenamefont
  {Torres}}]{PhysRevLett.115.220501}%
  \BibitemOpen
  \bibfield  {author} {\bibinfo {author} {\bibfnamefont {J.}~\bibnamefont
  {Svozil\'{\i}k}}, \bibinfo {author} {\bibfnamefont {A.}~\bibnamefont
  {Vall\'es}}, \bibinfo {author} {\bibfnamefont {J.}~\bibnamefont
  {Pe\ifmmode~\check{r}\else \v{r}\fi{}ina}}, \ and\ \bibinfo {author}
  {\bibfnamefont {J.~P.}\ \bibnamefont {Torres}},\ }\href {\doibase
  10.1103/PhysRevLett.115.220501} {\bibfield  {journal} {\bibinfo  {journal}
  {Phys. Rev. Lett.}\ }\textbf {\bibinfo {volume} {115}},\ \bibinfo {pages}
  {220501} (\bibinfo {year} {2015})}\BibitemShut {NoStop}%
\bibitem [{\citenamefont {Streltsov}\ \emph {et~al.}(2015)\citenamefont
  {Streltsov}, \citenamefont {Singh}, \citenamefont {Dhar}, \citenamefont
  {Bera},\ and\ \citenamefont {Adesso}}]{PhysRevLett.115.020403}%
  \BibitemOpen
  \bibfield  {author} {\bibinfo {author} {\bibfnamefont {A.}~\bibnamefont
  {Streltsov}}, \bibinfo {author} {\bibfnamefont {U.}~\bibnamefont {Singh}},
  \bibinfo {author} {\bibfnamefont {H.~S.}\ \bibnamefont {Dhar}}, \bibinfo
  {author} {\bibfnamefont {M.~N.}\ \bibnamefont {Bera}}, \ and\ \bibinfo
  {author} {\bibfnamefont {G.}~\bibnamefont {Adesso}},\ }\href {\doibase
  10.1103/PhysRevLett.115.020403} {\bibfield  {journal} {\bibinfo  {journal}
  {Phys. Rev. Lett.}\ }\textbf {\bibinfo {volume} {115}},\ \bibinfo {pages}
  {020403} (\bibinfo {year} {2015})}\BibitemShut {NoStop}%
\bibitem [{\citenamefont {Zhu}\ \emph {et~al.}(2017)\citenamefont {Zhu},
  \citenamefont {Ma}, \citenamefont {Cao}, \citenamefont {Fei},\ and\
  \citenamefont {Vedral}}]{PhysRevA.96.032316}%
  \BibitemOpen
  \bibfield  {author} {\bibinfo {author} {\bibfnamefont {H.}~\bibnamefont
  {Zhu}}, \bibinfo {author} {\bibfnamefont {Z.}~\bibnamefont {Ma}}, \bibinfo
  {author} {\bibfnamefont {Z.}~\bibnamefont {Cao}}, \bibinfo {author}
  {\bibfnamefont {S.-M.}\ \bibnamefont {Fei}}, \ and\ \bibinfo {author}
  {\bibfnamefont {V.}~\bibnamefont {Vedral}},\ }\href {\doibase
  10.1103/PhysRevA.96.032316} {\bibfield  {journal} {\bibinfo  {journal} {Phys.
  Rev. A}\ }\textbf {\bibinfo {volume} {96}},\ \bibinfo {pages} {032316}
  (\bibinfo {year} {2017})}\BibitemShut {NoStop}%
\bibitem [{\citenamefont {Tan}\ \emph {et~al.}(2018)\citenamefont {Tan},
  \citenamefont {Choi}, \citenamefont {Kwon},\ and\ \citenamefont
  {Jeong}}]{PhysRevA.97.052304}%
  \BibitemOpen
  \bibfield  {author} {\bibinfo {author} {\bibfnamefont {K.~C.}\ \bibnamefont
  {Tan}}, \bibinfo {author} {\bibfnamefont {S.}~\bibnamefont {Choi}}, \bibinfo
  {author} {\bibfnamefont {H.}~\bibnamefont {Kwon}}, \ and\ \bibinfo {author}
  {\bibfnamefont {H.}~\bibnamefont {Jeong}},\ }\href {\doibase
  10.1103/PhysRevA.97.052304} {\bibfield  {journal} {\bibinfo  {journal} {Phys.
  Rev. A}\ }\textbf {\bibinfo {volume} {97}},\ \bibinfo {pages} {052304}
  (\bibinfo {year} {2018})}\BibitemShut {NoStop}%
\bibitem [{\citenamefont {\ifmmode~\check{C}\else \v{C}\fi{}ernoch}\ \emph
  {et~al.}(2018)\citenamefont {\ifmmode~\check{C}\else \v{C}\fi{}ernoch},
  \citenamefont {Bartkiewicz}, \citenamefont {Lemr},\ and\ \citenamefont
  {Soubusta}}]{PhysRevA.97.042305}%
  \BibitemOpen
  \bibfield  {author} {\bibinfo {author} {\bibfnamefont {A.}~\bibnamefont
  {\ifmmode~\check{C}\else \v{C}\fi{}ernoch}}, \bibinfo {author} {\bibfnamefont
  {K.}~\bibnamefont {Bartkiewicz}}, \bibinfo {author} {\bibfnamefont
  {K.}~\bibnamefont {Lemr}}, \ and\ \bibinfo {author} {\bibfnamefont
  {J.}~\bibnamefont {Soubusta}},\ }\href {\doibase 10.1103/PhysRevA.97.042305}
  {\bibfield  {journal} {\bibinfo  {journal} {Phys. Rev. A}\ }\textbf {\bibinfo
  {volume} {97}},\ \bibinfo {pages} {042305} (\bibinfo {year}
  {2018})}\BibitemShut {NoStop}%
\bibitem [{\citenamefont {Fan}\ \emph {et~al.}(2019)\citenamefont {Fan},
  \citenamefont {Sun}, \citenamefont {Ding}, \citenamefont {Ming},
  \citenamefont {Yang}, \citenamefont {Wang},\ and\ \citenamefont
  {Ye}}]{Fan_2019}%
  \BibitemOpen
  \bibfield  {author} {\bibinfo {author} {\bibfnamefont {X.-G.}\ \bibnamefont
  {Fan}}, \bibinfo {author} {\bibfnamefont {W.-Y.}\ \bibnamefont {Sun}},
  \bibinfo {author} {\bibfnamefont {Z.-Y.}\ \bibnamefont {Ding}}, \bibinfo
  {author} {\bibfnamefont {F.}~\bibnamefont {Ming}}, \bibinfo {author}
  {\bibfnamefont {H.}~\bibnamefont {Yang}}, \bibinfo {author} {\bibfnamefont
  {D.}~\bibnamefont {Wang}}, \ and\ \bibinfo {author} {\bibfnamefont
  {L.}~\bibnamefont {Ye}},\ }\href {\doibase 10.1088/1367-2630/ab41b1}
  {\bibfield  {journal} {\bibinfo  {journal} {New J. Phys.}\ }\textbf {\bibinfo
  {volume} {21}},\ \bibinfo {pages} {093053} (\bibinfo {year}
  {2019})}\BibitemShut {NoStop}%
\bibitem [{\citenamefont {Theurer}\ \emph {et~al.}(2020)\citenamefont
  {Theurer}, \citenamefont {Satyajit},\ and\ \citenamefont
  {Plenio}}]{PhysRevLett.125.130401}%
  \BibitemOpen
  \bibfield  {author} {\bibinfo {author} {\bibfnamefont {T.}~\bibnamefont
  {Theurer}}, \bibinfo {author} {\bibfnamefont {S.}~\bibnamefont {Satyajit}}, \
  and\ \bibinfo {author} {\bibfnamefont {M.~B.}\ \bibnamefont {Plenio}},\
  }\href {\doibase 10.1103/PhysRevLett.125.130401} {\bibfield  {journal}
  {\bibinfo  {journal} {Phys. Rev. Lett.}\ }\textbf {\bibinfo {volume} {125}},\
  \bibinfo {pages} {130401} (\bibinfo {year} {2020})}\BibitemShut {NoStop}%
\bibitem [{\citenamefont {Ming}\ \emph {et~al.}(2021)\citenamefont {Ming},
  \citenamefont {Wang}, \citenamefont {Li}, \citenamefont {Fan}, \citenamefont
  {Song}, \citenamefont {Ye},\ and\ \citenamefont
  {Chen}}]{https://doi.org/10.1002/qute.202100036}%
  \BibitemOpen
  \bibfield  {author} {\bibinfo {author} {\bibfnamefont {F.}~\bibnamefont
  {Ming}}, \bibinfo {author} {\bibfnamefont {D.}~\bibnamefont {Wang}}, \bibinfo
  {author} {\bibfnamefont {L.-J.}\ \bibnamefont {Li}}, \bibinfo {author}
  {\bibfnamefont {X.-G.}\ \bibnamefont {Fan}}, \bibinfo {author} {\bibfnamefont
  {X.-K.}\ \bibnamefont {Song}}, \bibinfo {author} {\bibfnamefont
  {L.}~\bibnamefont {Ye}}, \ and\ \bibinfo {author} {\bibfnamefont {J.-L.}\
  \bibnamefont {Chen}},\ }\href {\doibase
  https://doi.org/10.1002/qute.202100036} {\bibfield  {journal} {\bibinfo
  {journal} {Adv. Quantum Technol.}\ }\textbf {\bibinfo {volume} {4}},\
  \bibinfo {pages} {2100036} (\bibinfo {year} {2021})}\BibitemShut {NoStop}%
\bibitem [{\citenamefont {Kumar}\ and\ \citenamefont
  {Dhar}(2016)}]{PhysRevA.93.062337}%
  \BibitemOpen
  \bibfield  {author} {\bibinfo {author} {\bibfnamefont {A.}~\bibnamefont
  {Kumar}}\ and\ \bibinfo {author} {\bibfnamefont {H.~S.}\ \bibnamefont
  {Dhar}},\ }\href {\doibase 10.1103/PhysRevA.93.062337} {\bibfield  {journal}
  {\bibinfo  {journal} {Phys. Rev. A}\ }\textbf {\bibinfo {volume} {93}},\
  \bibinfo {pages} {062337} (\bibinfo {year} {2016})}\BibitemShut {NoStop}%
\bibitem [{\citenamefont {Kalaga}\ \emph {et~al.}(2018)\citenamefont {Kalaga},
  \citenamefont {Leo\ifmmode~\acute{n}\else \'{n}\fi{}ski},\ and\ \citenamefont
  {Pe\ifmmode~\check{r}\else \v{r}\fi{}ina}}]{PhysRevA.97.042110}%
  \BibitemOpen
  \bibfield  {author} {\bibinfo {author} {\bibfnamefont {J.~K.}\ \bibnamefont
  {Kalaga}}, \bibinfo {author} {\bibfnamefont {W.}~\bibnamefont
  {Leo\ifmmode~\acute{n}\else \'{n}\fi{}ski}}, \ and\ \bibinfo {author}
  {\bibfnamefont {J.}~\bibnamefont {Pe\ifmmode~\check{r}\else \v{r}\fi{}ina}},\
  }\href {\doibase 10.1103/PhysRevA.97.042110} {\bibfield  {journal} {\bibinfo
  {journal} {Phys. Rev. A}\ }\textbf {\bibinfo {volume} {97}},\ \bibinfo
  {pages} {042110} (\bibinfo {year} {2018})}\BibitemShut {NoStop}%
\bibitem [{\citenamefont {Paul}\ and\ \citenamefont
  {Mukherjee}(2020)}]{PhysRevA.102.052209}%
  \BibitemOpen
  \bibfield  {author} {\bibinfo {author} {\bibfnamefont {B.}~\bibnamefont
  {Paul}}\ and\ \bibinfo {author} {\bibfnamefont {K.}~\bibnamefont
  {Mukherjee}},\ }\href {\doibase 10.1103/PhysRevA.102.052209} {\bibfield
  {journal} {\bibinfo  {journal} {Phys. Rev. A}\ }\textbf {\bibinfo {volume}
  {102}},\ \bibinfo {pages} {052209} (\bibinfo {year} {2020})}\BibitemShut
  {NoStop}%
\bibitem [{\citenamefont {Dai}\ \emph {et~al.}(2022)\citenamefont {Dai},
  \citenamefont {Fan},\ and\ \citenamefont {Qiu}}]{PhysRevA.105.022425}%
  \BibitemOpen
  \bibfield  {author} {\bibinfo {author} {\bibfnamefont {T.-Z.}\ \bibnamefont
  {Dai}}, \bibinfo {author} {\bibfnamefont {Y.}~\bibnamefont {Fan}}, \ and\
  \bibinfo {author} {\bibfnamefont {L.}~\bibnamefont {Qiu}},\ }\href {\doibase
  10.1103/PhysRevA.105.022425} {\bibfield  {journal} {\bibinfo  {journal}
  {Phys. Rev. A}\ }\textbf {\bibinfo {volume} {105}},\ \bibinfo {pages}
  {022425} (\bibinfo {year} {2022})}\BibitemShut {NoStop}%
\bibitem [{\citenamefont {Kalaga}\ and\ \citenamefont
  {Leo{\'n}ski}(2017)}]{kalaga2017quantum}%
  \BibitemOpen
  \bibfield  {author} {\bibinfo {author} {\bibfnamefont {J.~K.}\ \bibnamefont
  {Kalaga}}\ and\ \bibinfo {author} {\bibfnamefont {W.}~\bibnamefont
  {Leo{\'n}ski}},\ }\href
  {https://link.springer.com/article/10.1007/s11128-017-1627-6} {\bibfield
  {journal} {\bibinfo  {journal} {Quantum Inf. Process.}\ }\textbf {\bibinfo
  {volume} {16}},\ \bibinfo {pages} {175} (\bibinfo {year} {2017})}\BibitemShut
  {NoStop}%
\bibitem [{\citenamefont {Kalaga}\ \emph {et~al.}(2022)\citenamefont {Kalaga},
  \citenamefont {Leo{\'n}ski}, \citenamefont {Szcz\c{e}{\'s}niak},\ and\
  \citenamefont {Pe{\v{r}}ina~Jr}}]{kalaga2022mixedness}%
  \BibitemOpen
  \bibfield  {author} {\bibinfo {author} {\bibfnamefont {J.~K.}\ \bibnamefont
  {Kalaga}}, \bibinfo {author} {\bibfnamefont {W.}~\bibnamefont {Leo{\'n}ski}},
  \bibinfo {author} {\bibfnamefont {R.}~\bibnamefont {Szcz\c{e}{\'s}niak}}, \
  and\ \bibinfo {author} {\bibfnamefont {J.}~\bibnamefont {Pe{\v{r}}ina~Jr}},\
  }\href {\doibase 10.3390/e24030324} {\bibfield  {journal} {\bibinfo
  {journal} {Entropy}\ }\textbf {\bibinfo {volume} {24}},\ \bibinfo {pages}
  {324} (\bibinfo {year} {2022})}\BibitemShut {NoStop}%
\bibitem [{\citenamefont {Bartkiewicz}\ \emph {et~al.}(2013)\citenamefont
  {Bartkiewicz}, \citenamefont {Horst}, \citenamefont {Lemr},\ and\
  \citenamefont {Miranowicz}}]{PhysRevA.88.052105}%
  \BibitemOpen
  \bibfield  {author} {\bibinfo {author} {\bibfnamefont {K.}~\bibnamefont
  {Bartkiewicz}}, \bibinfo {author} {\bibfnamefont {B.}~\bibnamefont {Horst}},
  \bibinfo {author} {\bibfnamefont {K.}~\bibnamefont {Lemr}}, \ and\ \bibinfo
  {author} {\bibfnamefont {A.}~\bibnamefont {Miranowicz}},\ }\href {\doibase
  10.1103/PhysRevA.88.052105} {\bibfield  {journal} {\bibinfo  {journal} {Phys.
  Rev. A}\ }\textbf {\bibinfo {volume} {88}},\ \bibinfo {pages} {052105}
  (\bibinfo {year} {2013})}\BibitemShut {NoStop}%
\bibitem [{\citenamefont {Horst}\ \emph {et~al.}(2013)\citenamefont {Horst},
  \citenamefont {Bartkiewicz},\ and\ \citenamefont
  {Miranowicz}}]{PhysRevA.87.042108}%
  \BibitemOpen
  \bibfield  {author} {\bibinfo {author} {\bibfnamefont {B.}~\bibnamefont
  {Horst}}, \bibinfo {author} {\bibfnamefont {K.}~\bibnamefont {Bartkiewicz}},
  \ and\ \bibinfo {author} {\bibfnamefont {A.}~\bibnamefont {Miranowicz}},\
  }\href {\doibase 10.1103/PhysRevA.87.042108} {\bibfield  {journal} {\bibinfo
  {journal} {Phys. Rev. A}\ }\textbf {\bibinfo {volume} {87}},\ \bibinfo
  {pages} {042108} (\bibinfo {year} {2013})}\BibitemShut {NoStop}%
\bibitem [{\citenamefont {Miranowicz}\ \emph {et~al.}(2015)\citenamefont
  {Miranowicz}, \citenamefont {Bartkiewicz}, \citenamefont {Pathak},
  \citenamefont {Pe\ifmmode~\check{r}\else \v{r}\fi{}ina}, \citenamefont
  {Chen},\ and\ \citenamefont {Nori}}]{PhysRevA.91.042309}%
  \BibitemOpen
  \bibfield  {author} {\bibinfo {author} {\bibfnamefont {A.}~\bibnamefont
  {Miranowicz}}, \bibinfo {author} {\bibfnamefont {K.}~\bibnamefont
  {Bartkiewicz}}, \bibinfo {author} {\bibfnamefont {A.}~\bibnamefont {Pathak}},
  \bibinfo {author} {\bibfnamefont {J.}~\bibnamefont {Pe\ifmmode~\check{r}\else
  \v{r}\fi{}ina}}, \bibinfo {author} {\bibfnamefont {Y.-N.}\ \bibnamefont
  {Chen}}, \ and\ \bibinfo {author} {\bibfnamefont {F.}~\bibnamefont {Nori}},\
  }\href {\doibase 10.1103/PhysRevA.91.042309} {\bibfield  {journal} {\bibinfo
  {journal} {Phys. Rev. A}\ }\textbf {\bibinfo {volume} {91}},\ \bibinfo
  {pages} {042309} (\bibinfo {year} {2015})}\BibitemShut {NoStop}%
\bibitem [{\citenamefont {D\"ur}\ \emph {et~al.}(1999)\citenamefont {D\"ur},
  \citenamefont {Cirac},\ and\ \citenamefont {Tarrach}}]{PhysRevLett.83.3562}%
  \BibitemOpen
  \bibfield  {author} {\bibinfo {author} {\bibfnamefont {W.}~\bibnamefont
  {D\"ur}}, \bibinfo {author} {\bibfnamefont {J.~I.}\ \bibnamefont {Cirac}}, \
  and\ \bibinfo {author} {\bibfnamefont {R.}~\bibnamefont {Tarrach}},\ }\href
  {\doibase 10.1103/PhysRevLett.83.3562} {\bibfield  {journal} {\bibinfo
  {journal} {Phys. Rev. Lett.}\ }\textbf {\bibinfo {volume} {83}},\ \bibinfo
  {pages} {3562} (\bibinfo {year} {1999})}\BibitemShut {NoStop}%
\bibitem [{\citenamefont {Lee}\ \emph {et~al.}(2003)\citenamefont {Lee},
  \citenamefont {Chi}, \citenamefont {Oh},\ and\ \citenamefont
  {Kim}}]{PhysRevA.68.062304}%
  \BibitemOpen
  \bibfield  {author} {\bibinfo {author} {\bibfnamefont {S.}~\bibnamefont
  {Lee}}, \bibinfo {author} {\bibfnamefont {D.~P.}\ \bibnamefont {Chi}},
  \bibinfo {author} {\bibfnamefont {S.~D.}\ \bibnamefont {Oh}}, \ and\ \bibinfo
  {author} {\bibfnamefont {J.}~\bibnamefont {Kim}},\ }\href {\doibase
  10.1103/PhysRevA.68.062304} {\bibfield  {journal} {\bibinfo  {journal} {Phys.
  Rev. A}\ }\textbf {\bibinfo {volume} {68}},\ \bibinfo {pages} {062304}
  (\bibinfo {year} {2003})}\BibitemShut {NoStop}%
\bibitem [{\citenamefont {Mandel}\ and\ \citenamefont
  {Wolf}(1995)}]{mandel_wolf_1995}%
  \BibitemOpen
  \bibfield  {author} {\bibinfo {author} {\bibfnamefont {L.}~\bibnamefont
  {Mandel}}\ and\ \bibinfo {author} {\bibfnamefont {E.}~\bibnamefont {Wolf}},\
  }\href@noop {} {\emph {\bibinfo {title} {Optical Coherence and Quantum
  Optics}}}\ (\bibinfo  {publisher} {Cambridge University Press},\ \bibinfo
  {year} {1995})\BibitemShut {NoStop}%
\bibitem [{\citenamefont {Quintino}\ \emph {et~al.}(2015)\citenamefont
  {Quintino}, \citenamefont {V\'ertesi}, \citenamefont {Cavalcanti},
  \citenamefont {Augusiak}, \citenamefont {Demianowicz}, \citenamefont
  {Ac\'{\i}n},\ and\ \citenamefont {Brunner}}]{PhysRevA.92.032107}%
  \BibitemOpen
  \bibfield  {author} {\bibinfo {author} {\bibfnamefont {M.~T.}\ \bibnamefont
  {Quintino}}, \bibinfo {author} {\bibfnamefont {T.}~\bibnamefont {V\'ertesi}},
  \bibinfo {author} {\bibfnamefont {D.}~\bibnamefont {Cavalcanti}}, \bibinfo
  {author} {\bibfnamefont {R.}~\bibnamefont {Augusiak}}, \bibinfo {author}
  {\bibfnamefont {M.}~\bibnamefont {Demianowicz}}, \bibinfo {author}
  {\bibfnamefont {A.}~\bibnamefont {Ac\'{\i}n}}, \ and\ \bibinfo {author}
  {\bibfnamefont {N.}~\bibnamefont {Brunner}},\ }\href {\doibase
  10.1103/PhysRevA.92.032107} {\bibfield  {journal} {\bibinfo  {journal} {Phys.
  Rev. A}\ }\textbf {\bibinfo {volume} {92}},\ \bibinfo {pages} {032107}
  (\bibinfo {year} {2015})}\BibitemShut {NoStop}%
\bibitem [{\citenamefont {Horodecki}\ \emph {et~al.}(1995)\citenamefont
  {Horodecki}, \citenamefont {Horodecki},\ and\ \citenamefont
  {Horodecki}}]{HORODECKI1995340}%
  \BibitemOpen
  \bibfield  {author} {\bibinfo {author} {\bibfnamefont {R.}~\bibnamefont
  {Horodecki}}, \bibinfo {author} {\bibfnamefont {P.}~\bibnamefont
  {Horodecki}}, \ and\ \bibinfo {author} {\bibfnamefont {M.}~\bibnamefont
  {Horodecki}},\ }\href {\doibase https://doi.org/10.1016/0375-9601(95)00214-N}
  {\bibfield  {journal} {\bibinfo  {journal} {Phys. Lett. A}\ }\textbf
  {\bibinfo {volume} {200}},\ \bibinfo {pages} {340} (\bibinfo {year}
  {1995})}\BibitemShut {NoStop}%
\bibitem [{\citenamefont {Sadhukhan}\ \emph {et~al.}(2015)\citenamefont
  {Sadhukhan}, \citenamefont {Roy}, \citenamefont {Rakshit}, \citenamefont
  {Sen(De)},\ and\ \citenamefont {Sen}}]{Sadhukhan_2015}%
  \BibitemOpen
  \bibfield  {author} {\bibinfo {author} {\bibfnamefont {D.}~\bibnamefont
  {Sadhukhan}}, \bibinfo {author} {\bibfnamefont {S.~S.}\ \bibnamefont {Roy}},
  \bibinfo {author} {\bibfnamefont {D.}~\bibnamefont {Rakshit}}, \bibinfo
  {author} {\bibfnamefont {A.}~\bibnamefont {Sen(De)}}, \ and\ \bibinfo
  {author} {\bibfnamefont {U.}~\bibnamefont {Sen}},\ }\href {\doibase
  10.1088/1367-2630/17/4/043013} {\bibfield  {journal} {\bibinfo  {journal}
  {New J. Phys.}\ }\textbf {\bibinfo {volume} {17}},\ \bibinfo {pages} {043013}
  (\bibinfo {year} {2015})}\BibitemShut {NoStop}%
\bibitem [{\citenamefont {Pandya}\ \emph {et~al.}(2016)\citenamefont {Pandya},
  \citenamefont {Misra},\ and\ \citenamefont
  {Chakrabarty}}]{PhysRevA.94.052126}%
  \BibitemOpen
  \bibfield  {author} {\bibinfo {author} {\bibfnamefont {P.}~\bibnamefont
  {Pandya}}, \bibinfo {author} {\bibfnamefont {A.}~\bibnamefont {Misra}}, \
  and\ \bibinfo {author} {\bibfnamefont {I.}~\bibnamefont {Chakrabarty}},\
  }\href {\doibase 10.1103/PhysRevA.94.052126} {\bibfield  {journal} {\bibinfo
  {journal} {Phys. Rev. A}\ }\textbf {\bibinfo {volume} {94}},\ \bibinfo
  {pages} {052126} (\bibinfo {year} {2016})}\BibitemShut {NoStop}%
\bibitem [{\citenamefont {Toner}\ and\ \citenamefont
  {Verstraete}(2006)}]{toner2006monogamy}%
  \BibitemOpen
  \bibfield  {author} {\bibinfo {author} {\bibfnamefont {B.}~\bibnamefont
  {Toner}}\ and\ \bibinfo {author} {\bibfnamefont {F.}~\bibnamefont
  {Verstraete}},\ }\href@noop {} {\enquote {\bibinfo {title} {Monogamy of bell
  correlations and tsirelson's bound},}\ } (\bibinfo {year} {2006}),\ \Eprint
  {http://arxiv.org/abs/quant-ph/0611001} {arXiv:quant-ph/0611001 [quant-ph]}
  \BibitemShut {NoStop}%
\bibitem [{\citenamefont {Niculescu}\ and\ \citenamefont
  {Persson}(2006)}]{niculescu2006convex}%
  \BibitemOpen
  \bibfield  {author} {\bibinfo {author} {\bibfnamefont {C.}~\bibnamefont
  {Niculescu}}\ and\ \bibinfo {author} {\bibfnamefont {L.-E.}\ \bibnamefont
  {Persson}},\ }\href@noop {} {\emph {\bibinfo {title} {Convex functions and
  their applications}}},\ Vol.~\bibinfo {volume} {23}\ (\bibinfo  {publisher}
  {Springer},\ \bibinfo {year} {2006})\BibitemShut {NoStop}%
\bibitem [{\citenamefont {Boyd}\ and\ \citenamefont
  {Vandenberghe}(2004)}]{boyd2004convex}%
  \BibitemOpen
  \bibfield  {author} {\bibinfo {author} {\bibfnamefont {S.}~\bibnamefont
  {Boyd}}\ and\ \bibinfo {author} {\bibfnamefont {L.}~\bibnamefont
  {Vandenberghe}},\ }\href@noop {} {\emph {\bibinfo {title} {Convex
  optimization}}}\ (\bibinfo  {publisher} {Cambridge university press},\
  \bibinfo {year} {2004})\BibitemShut {NoStop}%
\bibitem [{\citenamefont {Bengtsson}\ and\ \citenamefont
  {{}\.{Z}yczkowski}(2017)}]{bengtsson_zyczkowski_2017}%
  \BibitemOpen
  \bibfield  {author} {\bibinfo {author} {\bibfnamefont {I.}~\bibnamefont
  {Bengtsson}}\ and\ \bibinfo {author} {\bibfnamefont {K.}~\bibnamefont
  {{}\.{Z}yczkowski}},\ }\href@noop {} {\emph {\bibinfo {title} {Geometry of
  Quantum States: An Introduction to Quantum Entanglement}}},\ \bibinfo
  {edition} {2nd}\ ed.\ (\bibinfo  {publisher} {Cambridge University Press},\
  \bibinfo {year} {2017})\BibitemShut {NoStop}%
\bibitem [{\citenamefont {Zyczkowski}\ and\ \citenamefont
  {Kus}(1994)}]{Zyczkowski_1994}%
  \BibitemOpen
  \bibfield  {author} {\bibinfo {author} {\bibfnamefont {K.}~\bibnamefont
  {Zyczkowski}}\ and\ \bibinfo {author} {\bibfnamefont {M.}~\bibnamefont
  {Kus}},\ }\href {\doibase 10.1088/0305-4470/27/12/028} {\bibfield  {journal}
  {\bibinfo  {journal} {J. Phys. A}\ }\textbf {\bibinfo {volume} {27}},\
  \bibinfo {pages} {4235} (\bibinfo {year} {1994})}\BibitemShut {NoStop}%
\bibitem [{\citenamefont {Cheng}\ and\ \citenamefont
  {Hall}(2017)}]{PhysRevLett.118.010401}%
  \BibitemOpen
  \bibfield  {author} {\bibinfo {author} {\bibfnamefont {S.}~\bibnamefont
  {Cheng}}\ and\ \bibinfo {author} {\bibfnamefont {M.~J.~W.}\ \bibnamefont
  {Hall}},\ }\href {\doibase 10.1103/PhysRevLett.118.010401} {\bibfield
  {journal} {\bibinfo  {journal} {Phys. Rev. Lett.}\ }\textbf {\bibinfo
  {volume} {118}},\ \bibinfo {pages} {010401} (\bibinfo {year}
  {2017})}\BibitemShut {NoStop}%
\end{thebibliography}

%

\end{document}